\documentclass[a4paper,11pt]{article} 
\usepackage{jheppub,amsthm,tikz,macros}
\usepackage{adjustbox}

\newcommand{\ho}{\mathrm{h}}
\newcommand{\co}[1][]{\underset{\left\lceil{#1}\right\rfloor}{\sum}}
\newcommand{\itk}[1]{
   \begin{tikzpicture}[baseline=(current bounding box.center),>={Triangle[open,width=0pt 20]},x=0.6cm,y=0.7cm]#1
   \end{tikzpicture}
   }
\newcommand{\act}{\blacktriangleright }
\newcommand{\ract}{\blacktriangleleft }
\newcommand{\cl}[1]{|#1\ra}
\newcommand{\cll}[1]{\la#1|}
\newcommand{\bx}{\boxtimes}
\usetikzlibrary{arrows.meta,cd}

\usepackage{bbm}
\newcommand{\1}{\mathbbm{1}}
\newcommand{\A}{\mathbbm{A}}

\usepackage{changes}

\usepackage{tensor}
\newcommand{\III}{\tensor*[_\1]{\1}{_\1} }
\newcommand{\IfI}{\tensor*[_\1]{f}{_\1} }
\newcommand{\AAI}{\tensor*[_\A]{\A}{_\1} }
\newcommand{\IAA}{\tensor*[_\1]{\A}{_\A} }
\newcommand{\AAA}{\tensor*[_\A]{\A}{_\A} }
\newcommand{\AfAA}{\tensor*[_\A]{f\A}{_\A} }
\newcommand{\AAV}{\tensor*[_{\A\ot \A}]{V}{_\1} }
\newcommand{\AAVf}{\tensor*[_{\A\ot \A}]{Vf}{_\1} }
\newcommand{\AAAAA}{\tensor*[_{\A\ot \A}]{\A}{} \ot \tensor*{\A}{_\A} }
\newcommand{\AAAAI}{\tensor*[_{\A\ot \A}]{\A}{} \ot \tensor*{\A}{_\1} }

\newcommand{\VAA}{\tensor*[_\1]{V}{_{\A\ot \A}} }

\begin{document}
\title{On a class of fusion 2-category symmetry: condensation completion of braided fusion category}
\author[a,b,c]{Wenjie Xi,}
\author[d]{Tian Lan,}
\author[d]{Longye Wang,}
\author[a]{Chenjie Wang,}
\author[b]{Wei-Qiang Chen}

\affiliation[a]{Department of Physics and HKU-UCAS Joint Institute for Theoretical and Computational Physics, The University of Hong Kong, Pokfulam Road, Hong Kong, China}

\affiliation[b]{Department of Physics and Shenzhen Key Laboratory of Advanced Quantum Functional Materials and Devices, Southern University of Science and Technology, Shenzhen 518055, China}

\affiliation[c]{Shenzhen Institute for Quantum Science and Engineering, and Department of Physics, Southern University of Science and Technology, Shenzhen, 518055, China}

\affiliation[d]{Department of Physics, The Chinese University of Hong Kong, Shatin, New Territories, Hong Kong, China}

\emailAdd{wjxi@connect.hku.hk}
\emailAdd{tlan@cuhk.edu.hk}
\emailAdd{wanglongye@cuhk.edu.hk}
\emailAdd{cjwang@hku.hk}
\emailAdd{chenwq@sustech.edu.cn}

\abstract{
Recently, many studies are focused on generalized global symmetry, a mixture of both invertible and non-invertible symmetries in various space-time dimensions. 
The complete structure of generalized global symmetry is described by higher fusion category theory. In this paper, we first review the construction of fusion 2-category symmetry $\Sigma \cB$ where $\cB$ is a a braided fusion category. In particular, we elaborate on the monoidal structure of $\Sigma \cB$ which determines fusion rules and controls the dynamics of topological operators/defects. We then take $\Sigma \mathrm{sVec}$ as an example to demonstrate how we calculate fusion rule, quantum dimension and 10j-symbol of the fusion 2-category. With our algorithm, all these data can be efficiently encoded and computed in computer program. The complete program has been uploaded to github\footnote{https://github.com/WJXI/2sVec.git}. Our work can be thought as explicitly computing the representation theory of $\cB$, in analogy to, for example the representation theory of $SU(2)$. The choice of basis bimodule maps are in analogy to the Clebsch-Gordon coefficients and the 10j-symbol are in analogy to the 6j-symbol.

}

\maketitle
\section{Introduction}

Symmetry serves as a guiding principle in physics. 
In modern language, a generalized symmetry~\cite{shao2023,VERLINDE1988,Bhardwaj2018,Chang2019,Choi2023_2,Inamura2021,thorngren2019,bhardwaj2022} is characterized by topological operators $U$ whose supports in space-time can undergo any smooth deformation. A $q$-form generalized symmetry in $d$-dimensional space-time is associated with topological operators that are supported on $d-q-1$ dimensional sub-manifolds. Typically, these topological operators form an algebra under composition.
There are primarily two classes of higher-form generalized symmetry: generalized global symmetry and generalized non-invertible symmetry.
For generalized global symmetry, the composition follows group multiplication and topological operators are unitary. By contrast, generalized non-invertible symmetry entails a more complicated composition structure, which is usually characterized by a fusion category. Moreover, the topological operators for generalized non-invertible symmetry may lack inverses.
Given a quantum system, its generalized symmetry is uniquely defined by the combination of all topological operators in various space-time dimensions. The corresponding mathematical structure is captured by a fusion higher category.

Fusion 1-category is widely applied in the study of many physical systems. In rational CFTs, non-invertible symmetries are generally 
characterized by Verlinde lines~\cite{VERLINDE1988,moore1989,Oshikawa1997,Frhlich2004}, for example, Kramers-Wannier duality~\cite{Kramers1941,Frhlich2004} line operator in Ising CFT. 
Mathematically, within a given rational CFT, each simple Verlinde line corresponds to a simple object in the fusion 1-category.
The properties of Verlinde lines, including composition, splitting, joining, re-coupling are encoded by the categorical data of the fusion category. These categorical data can be explicitly expressed in terms of the fusion rules $N_{ab}^c$ and the F symbol.
(Braided) Fusion 1-category has also been used to study other physical systems such as 2+1D topological order where F symbol is cornerstone for constructing lattice model and $N_{ab}^c$ determines fusion structure of anyons. For fusion 1-category,  there are many ways to obtain the F symbol, including field theory~\cite{kawagoe2021anomalies}, representation theory~\cite{wigner2012group} (where the 6j symbol of $SU(2)$ is an example), or directly solving the pentagon equation~\cite{colazzo2020set,mazzotta2023idempotent}. However, calculating the F symbol is generally a very intricate process.

Recently, in the high-energy physics community, there has been a growing interest in studying the properties of higher-form non-invertible symmetries in higher space-time using various approaches~\cite{shao2023,Choi2023,choi2022,Damia2023,kong2014,choi2022_2,Kaidi2022,Choi2023_2,copetti2023,Antinucci2022,Bhardwaj2023,bhardwaj2022,bartsch2022}. For instance, some studies are focused on simplest non-invertible symmetry associated with the Kramers-Wannier duality in higher dimensions~\cite{choi2022_2,Kaidi2022}.
It has been realized that fusion 2-category plays an important rule in studying higher-form (non-invertible) symmetry~\cite{xi2021,inamura2023,delcamp2023,kong2014,Antinucci2022,Bhardwaj2023}, however, the study of higher fusion category is still in its infancy.
Even the rigorous definition of fusion 2-category is only proposed in 2018~\cite{douglas2018}, and few examples of fusion 2-category have been explicitly constructed.  By now, the only examples of fusion 2-category that we can list all the explicit data, are of the form $2\mathrm{Vec}_G^w$.
Moreover, approaches to finding examples with complete categorical data are not widely explored.
For many proposed QFTs or lattice models with higher form symmetries, determining their complete generalized global symmetries or identifying the fusion higher categories that should be used to characterize them remains unclear. 
Even for some systems that we can find all (higher-form) symmetries, the categorical data is often incomplete.

An important ingredient of fusion 2-category is the 10j-symbol~\cite{douglas2018},\footnote{In Walker-Wang model, the 10j-symbol of a presemisimple 2-category, which is the delooping of a unitary braided fusion category, is provided.} which is analogous to the F symbol of fusion 1-category. However, despite its significance, it has not received thorough investigation in current literature focusing on higher symmetries.
Currently, the only known examples of the 10j-symbols are the 4-cocycles of fusion 2-categories such as $2\mathrm{Vec}_G^\omega$, where only invertible objects are exist. Practically, attempts to directly solve the hexagon equation to obtain the 10j-symbol for arbitrary fusion 2-categories face formidable challenges. Even for numerical calculation, the computational cost is incredibly high. Therefore, it is highly demanded to find a practical way to obtain the complete categorical data for fusion 2-category, and derive a few simple but non-trivial examples. 

The explicit categorical data is also very important for studying physical systems with generalized global symmetries. For example, the data of fusion 2-category can be used to construct lattice model of 3+1D topological order~\cite{xi2021} and its boundary~\cite{inamura2023}. Since the data of fusion 1-category has been used to characterize 1+1D CFT, hopefully, we may use fusion 2-category to study 2+1D CFT which is also closely related to quantum phase transition in 2+1D.

In this paper, we propose an algorithm for systematically constructing examples for a class of fusion 2-category $\Sigma \cB$, the condensation completion of a braided fusion 1-category $\cB$, and obtaining all its categorical data.
As a preliminary application, we compute the full data for $\Sigma \mathrm{sVec}$.
Roughly speaking, we give the coefficients for all possible kinematics, including fusing, bending, braiding, recoupling, etc., of fermions and open Majorana chains.

The paper is organized as the following.
In section~\ref{pre}, we first review the construction of the braided fusion 1-category $\cB$. We also review separable algebras in $\cB$, bimodules of separable algebras and bimodule maps between the bimodules. we then review the construction of the fusion 2-category $\Sigma \cB$.
In section~\ref{GenMono}, we elaborate the monoidal structure of $\Sigma \cB$, which is mainly consist of fusion algebra of objects and 1-morphisms, which correspond to topological operators/defects, and 10j-symbol which captures the generalized crossing relations between the operators/defects.
In section~\ref{sVecspher}, we impose spherical condition for $\Sigma \cB$ which gives each topological defect a quantum dimension, a pairing structure of section and retraction bimodule maps and determines the normalization factor of the 10j-symbols.
In section~\ref{sVecbasic}, we explicitly compute the objects, 1-morphisms, 2-morphisms, fusion algebra and quantum dimension of a simple but fundamentally important example: $\Sigma \mathrm{sVec}$. 
In section~\ref{sVec10j}, we write down the explicit form of 10j-symbol with a chosen base. All the 10j-symbols of $\Sigma \mathrm{sVec}$ and the complete computer program has been uploaded to github. With our algorithm, all the categorical data of $\Sigma \cB$ can be efficiently computed with computer program.

\section{Preliminaries}\label{pre}
\subsection{Braided fusion category $\cB$}
Here we only introduce the properties of a braided fusion category $(\cB,\ot,\one,\alpha,c)$ that are relevant to this paper. For concrete and detailed definition, please see for example the textbook~\cite{EGNO}.
\begin{definition}
    [Monoidal category]A monoidal category $(\cB, \ot, \one, \alpha)$ is a category $\cB$ equipped with a monoidal structure consists of
    \begin{itemize}
        \item a tensor product $\ot: \cB \times \cB \rightarrow \cB$,
        \item a tensor unit $\one$ with $\one\ot X =  X = X \ot \one, \forall X \in \cB$.
        \item an associator $\alpha$, i.e. natural isomorphisms $\alpha_{X, Y, Z}: (X\ot Y)\ot Z \rightarrow X \ot (Y\ot Z) $ that satisfy the pentagon diagrams.
    \end{itemize}
\end{definition}
\begin{definition}
    [Fusion category]A fusion category $(\cB,\ot,\one,\alpha)$ is a category satisfies the following conditions:
    \begin{itemize}
        \item $(\cB,\ot,\one,\alpha)$ is a monoidal category,
        \item $\cB$ is $\C$-linear,
        \item $\cB$ is rigid,
        \item $\cB$ is finite semi-simple,
        \item the tensor unit $\one$ is a simple object.
    \end{itemize}
\end{definition}
\begin{definition}
    [Braided fusion category] A braided fusion category $(\cB,\ot,\one,\alpha, c)$ is a fusion category $(\cB,\ot,\one,\alpha)$ equipped with a braiding $c$, i.e. natural isomorphisms $c_{X,Y}:X\ot Y\xrightarrow{\sim}Y\ot X$ that satisfy the hexagon diagrams.
\end{definition}
\begin{remark}
    For concreteness, we restrict to the case where the objects in $\cB$ are ``vector spaces'' with certain structures, while the morphisms in $\cB$ are ``linear maps'' preserving the structures. Some examples include representation categories of groups or quasi-triangular Hopf algebras, and pointed braided fusion categories (i.e. finite pre-metric groups).
\end{remark}

\subsection{Algebras and modules in a braided fusion category}
Given a braided fusion category $(\cB,\ot,\one,\alpha,c)$, the algebras and its modules in $\cB$ are defined as the following.
\begin{definition}
    [Algebra in a braided fusion category] An algebra is a pair $(A,\ m:A\ot A\to A)$, where $A$ is an object in $\cB$ and the multiplication morphism $A\ot A \xrightarrow{m} A$ satisfies the following diagram
    \begin{equation}
        \begin{tikzcd}
        	{(A\ot A)\ot A} && {A\ot(A\ot A)} \\
        	{A\ot A} && {A\ot A} \\
        	& A
        	\arrow["{\alpha_{A,A,A}}", from=1-1, to=1-3]
        	\arrow["{m\ot\id_A}"', from=1-1, to=2-1]
        	\arrow["{\id_A\ot m}", from=1-3, to=2-3]
        	\arrow["m"', from=2-1, to=3-2]
        	\arrow["m", from=2-3, to=3-2]
        \end{tikzcd}.
    \end{equation}  
    The algebra may be denoted as $A$ for simplicity.
\end{definition}

\begin{example}
    $(\one, m=\id_\one)$ is the trivial algebra in $\cB$, and will be denoted as $\1$ in the paper.
\end{example}

\begin{definition}
    [Right $A$-module] Given an algebra $A$. A right $A$-module is a pair $(M,\ r:M \ot A\to M)$, where $M$ is an object in $\cB$ and $r$ is a morphism $M_A\ot A\to M_A$ such that the following diagram commutes
    \begin{equation}
        \begin{tikzcd}
        	{(M\ot A)\ot A} && {M\ot(A\ot A)} \\
        	{M\ot A} && {M\ot A} \\
        	& M
        	\arrow["{\alpha_{M,A,A}}", from=1-1, to=1-3]
        	\arrow["{r\ot\id_A}"', from=1-1, to=2-1]
        	\arrow["{\id_{M}\ot m}", from=1-3, to=2-3]
        	\arrow["r"', from=2-1, to=3-2]
        	\arrow["r", from=2-3, to=3-2]
        \end{tikzcd}
    \end{equation}
\end{definition}
\begin{remark}
    A left $A$-module $(N ,\ l:A\ot N\to N)$ is defined in the same way but with a left action $l$.
\end{remark}

\begin{definition}
    [$B$-$A$-Bimodule] Given two algebras $A$ and $B$. A $B$-$A$-bimodule is a triple $(M,\ l:B\ot M\to M,\ r:M\ot A\to M)$, where $(M,\ l)$ is a left $B$-module, $(M,\ r)$ is a right $A$-module, and that the following diagram commutes
    \begin{equation}
        \begin{tikzcd}
        	{(B\ot M)\ot A} && {B\ot(M\ot A)} \\
        	{M\ot A} && B \ot M \\
        	& M
        	\arrow["{\alpha_{B,M,A}}", from=1-1, to=1-3]
        	\arrow["{l\ot\id_A}"', from=1-1, to=2-1]
        	\arrow["{\id_B\ot r}", from=1-3, to=2-3]
        	\arrow["r"', from=2-1, to=3-2]
        	\arrow["l", from=2-3, to=3-2]
        \end{tikzcd}
    \end{equation}
\end{definition}

\begin{remark}
In the following, for simplicity, we will denote a $B$-$A$-bimodule $(M,\ l,\ r)$ with $\tensor*[_B]{M}{_A}$ or just the object $M$, if their meanings are evident from the context.  It also works for the left and right modules.
\end{remark}

\begin{remark}
    A left $A$-module $\tensor*[_A]{M}{}$ can be regarded as an $A$-$\1$-bimodule $\tensor*[_A]{M}{_\1}$, while a right $A$-module $\tensor*{N}{_A}$ can be regarded as a $\1$-$A$-bimodule $\tensor*[_\1]{N}{_A}$.
\end{remark}

\begin{example}
Given an algebra $(A, m)$ in $\cB$.
\begin{itemize}
    \item $\tensor*[_A]{A}{_A} \equiv (A, m, m)$ is an $A$-$A$-bimodule.
    \item $\tensor*[_A]{A}{} \ot \tensor*{A}{_A} \equiv (A\ot A,\ l_A,\ r_A)$ is an $A$-$A$-bimodule, where $l_A,\ r_A$ are defined as
    \begin{equation}
        \begin{tikzcd}
        	{l_A:A\ot(A\ot A)} & {(A\ot A)\ot A} & {A\ot A}
        	\arrow["{\alpha_{A,A,A}^{-1}}", from=1-1, to=1-2]
        	\arrow["{m\ot \id_A}", from=1-2, to=1-3]
        \end{tikzcd}
    \end{equation}
    \begin{equation}
        \begin{tikzcd}
        	{r_A:(A\ot A)\ot A} & {A\ot (A\ot A)} & {A\ot A}
        	\arrow["{\alpha_{A,A,A}}", from=1-1, to=1-2]
        	\arrow["{\id_A\ot m}", from=1-2, to=1-3]
        \end{tikzcd}
    \end{equation}
\end{itemize}
\end{example}

\begin{example}
    Given a $C$-$B$-bimodule $(M,\ l_M,\ r_M)$ and a $B$-$A$-bimodule $(N,\ l_N,\ r_N)$.
    \begin{itemize}
        \item The triple $(M\ot N,\ l_{MN},\ r_{MN})$ is a $C$-$A$-bimodule where $l$,$r$ are defined as
            \begin{equation}
                \begin{tikzcd}
                	{l_{MN}:C\ot(M\ot N)} & {(C\ot M)\ot N} & {M\ot N}
                	\arrow["{\alpha^{-1}}", from=1-1, to=1-2]
                	\arrow["{l_M\ot \id_N}", from=1-2, to=1-3]
                \end{tikzcd}
            \end{equation}
            \begin{equation}
                \begin{tikzcd}
                	{r_{MN}:(M\ot N)\ot A} & {M\ot (N\ot A)} & {M\ot N}
                	\arrow["{\alpha}", from=1-1, to=1-2]
                	\arrow["{\id_M\ot r_N}", from=1-2, to=1-3]
                \end{tikzcd}
            \end{equation}
        \item The triple $(M\ot B\ot N,\ l_{MBN},\ r_{MBN})$ is a $C$-$A$-bimodule where $l_{MBN},r_{MBN}$ is defined as
            \begin{equation}
                \begin{tikzcd}
                	{l_{MBN}:C\ot(M\ot B\ot N)} & {(C\ot M)\ot B\ot N} && M\ot B\ot N
                	\arrow["\alpha^{-1}", from=1-1, to=1-2]
                	\arrow["l_M\ot\id_B\ot\id_N", from=1-2, to=1-4]
                \end{tikzcd}
            \end{equation}
            \begin{equation}
                \begin{tikzcd}
                	{r_{MBN}:(M\ot B\ot N)\ot A} & {M\ot B\ot (N\ot A)} && M\ot B\ot N
                	\arrow["\alpha", from=1-1, to=1-2]
                	\arrow["\id_M\ot\id_B\ot r_{N}", from=1-2, to=1-4]
                \end{tikzcd}
            \end{equation}

    \end{itemize}
\end{example}
\begin{definition}
    [Module map] Given two right $A$-modules $(M,\ r_M:M\ot A\to M)$ and $(N,\ r_N:N\ot A\to N)$. A right $A$-module map is a morphism $f:M\to N$ in $\cB$ such that the following diagram commutes
    \begin{equation}
        \begin{tikzcd}
        	{M\ot A} && M \\
        	\\
        	{N\ot A} && N
        	\arrow["{r_M}", from=1-1, to=1-3]
        	\arrow["{f\ot \id_A}"', from=1-1, to=3-1]
        	\arrow["{r_N}"', from=3-1, to=3-3]
        	\arrow["f", from=1-3, to=3-3]
        \end{tikzcd}
    \end{equation}
\end{definition}

\begin{definition}
    [Bimodule map] Given two $B$-$A$-bimodules $(M,\ l_M:B\ot M\to M,\ r_M:M\ot A\to M)$, $(N,\ l_N:B\ot N\to N,\ r_N:N\ot A\to N)$. A bimodule map is a morphism $f:M\to N$ such that $f$ is both a left module map and a right module map.
\end{definition}
\begin{definition}
    [Separable algebra] An algebra $(A,\ m:A\ot A\to A)$ is called separable if $m:A\ot A\to A$ admits an $A$-$A$-bimodule map $\sigma:A\to A\ot A$ such that the composition $m\circ\sigma=\id_A$.
\end{definition}

\begin{definition}
    [Relative tensor product of bimodules]
    \label{rel_tensor_prod}
     Given a $C$-$B$-bimodule $(M,\ l_M:C\ot M\to M,\ r_M:M\ot B\to M)$ and a $B$-$A$-bimodule $(N,\ l_N:B\ot N\to N,\ r_N:N\ot A\to N)$. The relative tensor product (${M\ot[B]N}$, $\pi$), or simply ${M\ot[B]N}$, in $\cB$ is the coequalizer shown below
    \begin{equation}
        \begin{tikzcd}
        	{M\ot B\ot N} & {M\ot N} & {M\ot[B]N}
        	\arrow["{r_M\ot \id_N}", shift left=1, from=1-1, to=1-2]
        	\arrow["{\id_M\ot l_N}"', shift right=1, from=1-1, to=1-2]
        	\arrow["\pi", from=1-2, to=1-3]
        \end{tikzcd}
    \end{equation}
\end{definition}

\begin{remark}
    The universal property of $M\ot[B]N$ is given by the following commuting diagram
    \begin{equation}
        \begin{tikzcd}
        	{M\ot B\ot N} & {M\ot N} & {M\ot[B] N} \\
        	&& X
        	\arrow["\pi", from=1-2, to=1-3]
        	\arrow["{r_M\ot \id_N}", shift left=1, from=1-1, to=1-2]
        	\arrow["{\id_M\ot l_N}"', shift right=1, from=1-1, to=1-2]
        	\arrow["\exists ! {\tilde{h}}", dashed, from=1-3, to=2-3]
        	\arrow[" h"', from=1-2, to=2-3]
        \end{tikzcd}
    \end{equation}
for any $X\in\cB$, where $h \circ {r_M\ot \id_N}= h \circ {\id_M\ot l_N}$.
\end{remark}

\begin{remark}
    Given a $C$-$B$-bimodule $(M,\ l_M,\ r_M)$ and a $B$-$A$-bimodule $(N,\ l_N,\ r_N)$. $M\ot[B]N$ is the cokernel of the map $f=r_M\ot \id_N-\id_M\ot l_N$ as following
    \begin{equation}
        \begin{tikzcd}
            {M\ot B\ot N} & {M\ot N} & {M\ot[B] N} \\
            && X
            \arrow["\pi", from=1-2, to=1-3]
            \arrow["{f}", shift left=1, from=1-1, to=1-2]
            \arrow["{0}"', shift right=1, from=1-1, to=1-2]
            \arrow["\exists!{\tilde{h}}", dashed, from=1-3, to=2-3]
            \arrow[" h"', from=1-2, to=2-3]
        \end{tikzcd}
    \end{equation}
, where $h \circ f= h \circ 0$.
\end{remark}
\begin{remark}
    The relative tensor product $M \ot[B] N$ is uniquely determined up to a canonical isomorphism.
\end{remark}

\begin{remark}
    For a left $A$-module $(N ,\ l:A\ot N\to N)$, in general, $A \ot[A] N \xrightarrow[\sim]{\tilde{l}} N$, where $\tilde{l}$ is given by the universal property of the relative tensor product as
    \begin{equation}
        \begin{tikzcd}
            {A\ot A\ot N} & {A\ot N} & {A\ot[A] N} \\
            && N
            \arrow["\pi", from=1-2, to=1-3]
            \arrow["m \ot id_N", shift left=1, from=1-1, to=1-2]
            \arrow["id_A \ot l"', shift right=1, from=1-1, to=1-2]
            \arrow["\tilde{l}","\sim"', dashed, from=1-3, to=2-3]
            \arrow["l"', from=1-2, to=2-3]
        \end{tikzcd}
    \end{equation}
\end{remark}

\begin{definition}
    [Relative tensor product of bimodule maps]
    \label{rel_tensor_prod_maps}
For any two bimodule maps $f:{}_C M {}_B\to {}_C M' {}_B$ and $g:{}_B N {}_A\to {}_B N' {}_A$, the relative tensor product $f\ot[B]g$ is given by the universal property of the relative tensor product of the bimodules
\begin{equation}
    \begin{tikzcd}
        {M\ot B\ot N} && M\ot N && M\ot[B] N \\
            \\
        M'\ot B\ot N' && M'\ot N' && M'\ot[B] N'
        \arrow["r_M\ot \id_N", shift left=2, from=1-1, to=1-3]
        \arrow["\id_M \ot l_N"', shift right=2, from=1-1, to=1-3]
        \arrow["\pi", from=1-3, to=1-5]
        \arrow["{f\ot[B]g}", dashed, from=1-5, to=3-5]
        \arrow["f\ot\mathrm{id}_B \ot g"', from=1-1, to=3-1]
        \arrow["r_{M'}\ot\id_{N'}", shift left=2, from=3-1, to=3-3]
        \arrow["{\pi'}"', from=3-3, to=3-5]
        \arrow["\id_{M'}\ot l_{N'}"', shift right=2, from=3-1, to=3-3]
        \arrow["f\ot g"', from=1-3, to=3-3]
    \end{tikzcd}
    \end{equation}
\end{definition}

\subsection{Fusion 2-category and 10j-symbol}

Here we introduce the definition of a fusion 2-category $\cC$.  We only include the properties that are relevant to our paper. For concrete and detailed definition, please see ,for example, the Ref \cite{douglas2018}.

\begin{definition}
    [Monoidal 2-category]A monoidal 2-category $\cC$ is a 2-category $\cC$ equipped with a monoidal structure consists of
    \begin{itemize}
        \item the objects $(A, B, \cdots)$, 1-morphisms $(f, g, \cdots)$, and 2-morphisms $(\alpha, \beta, \cdots)$,
        \item the hom space $\Hom(A,B)$, which is a 1-category, consists of all 1-morphisms from object $A$ to object $B$ and the 2-morphisms between these 1-morphisms.
        \item the composition functor $\circ$
        \begin{align*}
            \circ: \Hom(A,B) \xt \Hom(B,C)  &\to \Hom(A,C),\\ \nonumber
                (f, g)&\mapsto g \circ f
        \end{align*}
        \item an associator 2-isomorphism
        \begin{align}
            \lambda_{f, g, h}: (f \circ g) \circ h \rightarrow f \circ (g \circ h)
        \end{align}
        for $f : C \to D $, $g : B \to C $, and $h: A \to B$
        \item a monoidal unit $\one$,
        \item a tensor product $\bx$, which are defined as 2-functors
        \begin{align}
            A \bx - : \cC \to \cC, &&
            - \bx A : \cC \to \cC
        \end{align}
        for each object $A \in \cC$,
        \item an interchange 2-isomorphism
        \begin{align}
            \phi_{f,g}:  (f \bx Z)\circ(B \bx g) \rightarrow (C\bx g)\circ (f \bx Y)
        \end{align}
        for each pair of 1-morphisms: $f : B \to C $ and $g : Y \to Z $,
        \item an invertible natural associativity 1-morphism 
        \begin{align}
            \Lambda_{A, B, C}: (A\bx B)\bx C \to A \bx (B\bx C)
        \end{align}
        for any objects $A, B, C \in \cC$, which tracks the associativity of the tensor product of objects,
        \item a pentagonator 2-isomorphism
        \begin{equation}
            \beta_{A,B,C,D}: (A\bx \Lambda_{B,C,D}) \circ \Lambda_{A,B\bx C,D} \circ (\Lambda_{A,B,C}\bx D) \rightarrow \Lambda_{A,B,C\bx D}\circ \Lambda_{A\bx B, C,D}
        \end{equation}
        for any objects $A, B, C, D \in \cC$.
    \end{itemize}
\end{definition}

\begin{definition}
    [Fusion 2-category]A fusion 2-category is a finite semisimple monoidal 2-category that has left and right duals for objects and a simple monoidal unit. 
\end{definition}

For given objects $A, B, C, K$ in $\cC$, the associativity 1-morphism $\Lambda_{A, B, C}$ will induce an equivalent functor
\begin{align}
\label{gen-asso-func}
   \Lambda_{A,B,C}\circ - : \Hom(K,(A\bx B)\bx C) \xrightarrow[]{} \Hom(K, A\bx (B\bx C)).
\end{align}
And for given objects $A, B, C, D, K$ in $\cC$, the pentagonator can induce a natural transformation between two equivalent functors as shown below
\begin{equation}
\label{gen-def-10j}
    \begin{tikzcd}
        \Hom(K,((A\bx B)\bx C)\bx D) \arrow[ddd,bend right=50,"\Lambda_{A,B,C\bx D}\circ \Lambda_{A\bx B, C,D}\circ -"{left},""{name=L,right}] 
        \arrow[ddd,bend left=50,"(A\bx \Lambda_{B,C,D}) \circ \Lambda_{A,B\bx C,D} \circ (\Lambda_{A,B,C}\bx D)\circ -"{right},""{name=R,left}]
        \\ \\ \\ \Hom(K,A\bx (B\bx (C\bx D))) \arrow[rightarrow, from=R, to=L, "\beta_{A,B,C,D}\circ -"'],
    \end{tikzcd}
\end{equation}
which is characterized by the 10j-symbol.

\subsection{The 2-category \texorpdfstring{$\Sigma\cB$}{ΣB}}
\label{sec:sigmaB}
In this paper, we will focus on fusion 2-category $\Sigma \cB$, the condensation completion of a braided fusion 1-category $\cB$.  We consider only the case where $\Sigma \cB$ has a spherical structure.  The definition of $\Sigma \cB$ as a 2-category is given below.  The monoidal structure and spherical structure of $\Sigma \cB$ will be discussed in Sec. \ref{GenMono} and Sec. \ref{sVecspher}.

\begin{definition}
    Given a braided fusion category $\cB$, its condensation completion~\cite{gaiotto2019cond,Kong_2022} $\Sigma\cB$, as a 2-category, consists of the following data.
    \begin{itemize}
        \item Objects are separable algebras in $\cB$.
        \item Given two objects $A,B$, the hom space $\Hom(A,B)$ is a 1-category consists of $B$-$A$-bimodules (as objects) and $B$-$A$-bimodule maps (as morphisms).
        \item The composition $\circ$ of hom spaces is given by the relative tensor product of bimodules and bimodule maps defined in Def.~\ref{rel_tensor_prod} and ~\ref{rel_tensor_prod_maps}
        \begin{align}
            \circ: \Hom(A,B) \xt \Hom(B,C) &\to \Hom(A,C),\\ \nonumber
            (_B N_A, {}_C M_B)&\mapsto M\circ N:={}_C M_B \ot[B] {}_B N_A,\\ \nonumber
            (g,f)&\mapsto f\circ g:=f\ot[B]g.
        \end{align}
    \end{itemize}
\end{definition}

\section{The monoidal structure of \texorpdfstring{$\Sigma\cB$}{ΣB}} \label{GenMono}

In this section, we describe the monoidal structure of $\Sigma\cB$ induced from the braided monoidal structure of $\cB$.

\subsection{Tensor product $\bx$ in $\Sigma\cB$}
\label{sec:ten_pro_bim}
The tensor product $\bx$ in $\Sigma\cB$ is induced by the tensor product $\ot$ in $\cB$.  Given two objects $A,B\in \Sigma\cB$, i.e. two separable algebras in $\cB$, $A\bx B:= (A \ot B, m_{A\ot B})$ with multiplication defined as
\begin{equation}
    \begin{tikzcd}
    (A\ot B)\ot (A \ot B)\arrow[r,"\alpha"] \arrow[dd,"m_{A\ot B}"]& A\ot(B\ot A)\ot B\arrow[d,rightarrow,"\id_A\ot c_{B,A}\ot \id_B"] \\
    &A\ot(A\ot B)\ot B\arrow[d,"\alpha"]\\
    A\ot B & (A\ot A)\ot (B\ot B)\arrow[l,"m_A\ot m_B"]
    \end{tikzcd} 
\end{equation}
is also a separable algebra in $\cB$, and hence an object in $\Sigma\cB$. 

Let $_C N_B $ and $ {}_Z P_Y$ be two 1-morphisms in $\Sigma\cB$, i.e. two bimodules in $\cB$. $N\ot P$ has a natural structure of $C\ot Z$-$B\ot Y$-bimodule, where the right module structure is defined as
\begin{equation}
    \begin{tikzcd}
    (N\ot P)\ot (B \ot Y)\arrow[r,"\alpha"] \arrow[dd,"r_{N \ot P}"]& N\ot(P\ot B)\ot Y\arrow[d,rightarrow,"\id_N\ot c_{P,B}\ot \id_Y"] \\
    &N\ot(B\ot P)\ot Y\arrow[d,"\alpha"]\\
    N\ot P & (N\ot B)\ot (P\ot Y)\arrow[l,"r_N\ot r_P"]
    \end{tikzcd}
\end{equation}
The left module structure can be defined similarly.  Then the tenor product of $N$ and $P$ in $\Sigma\cB$ is defined as $_{C\bx Z}{N \bx P}_{B\bx Y} := N \ot P$.  For the tensor product of 2-morphisms $f$ and $g$ in $\Sigma \cB$, since $f \ot g$ is automatically a bimodule map in $\cB$, we have $f \bx g := f \ot g$.  

Therefore, we do not distinguish the tensor product $\bx$ in $\Sigma \cB$ and $\ot$ in $\cB$ in the following.

\subsection{Interchange law}
The tensor product must be compatible with the bimodule composition $\circ$, which means $(M\ot P)\circ(N\ot Q)$ must be equivalent to $(M\circ N)\ot (P\circ Q)$.  This can be satisfied by the 2-isomorphism 
\begin{align}
(M\ot P)\circ(N\ot Q)  \stackrel{\tl{c}_{P,N;M,Q}} {\longrightarrow} (M\circ N)\ot (P\circ Q),
\end{align}
for the bimodules $_D M_C, {}_C N_B, {}_Z P_Y, {}_Y Q_X$.  $\tilde{c}_{P,N;M,Q}$ is induced by the braiding $c_{P, N}$ in $\cB$ via the universal property of the relative tensor product
\begin{equation}
\label{eq:ctilde}
    \begin{tikzcd}
        M\ot P\ot N\ot Q  \arrow[r,rightarrow,"\id\ot c_{P,N}\ot \id"]
        \arrow[d,"{\ot[C \ot Y]}"]
        & M \ot N\ot P \ot Q
        \arrow[d,"{\ot[C]\ \ot \ \ot[Y]}"]
        \\
        (M\ot P)\circ(N\ot Q) \arrow[r,rightarrow,"\tl{c}_{P,N;M, Q}"]
        &(M\circ N)\ot (P\circ Q)
    \end{tikzcd}
\end{equation}
where the associator $\alpha$ has been dropped for simplicity.

Then the interchanger $\phi_{N, P}$ for bimodules $N\in \Hom(B,C)$ and $P\in \Hom(Y,Z)$ is given by
\begin{align}
\phi_{N, P}: &(N \ot Z)\circ(B\ot P)\stackrel{\tl{c}_{Z,B;N, P}}{\longrightarrow} (N\circ B)\ot (Z\circ P) \xrightarrow{\tilde{r}_N \ot \tilde{l}_P} N\ot P \nonumber\\
& \xrightarrow{\tilde{l}^{-1}_N \ot \tilde{r}^{-1}_P} (C\circ N)\ot(P\circ Y) \stackrel{\tl{c}_{P,N;C, Y}^{-1}}{\longrightarrow}  (C\ot P)\circ (N\ot Y).\label{eq.interchange}
\end{align}
In the following, we will denote $\tilde{c}_{P,N;M,Q}$ as $\tilde{c}_{P,N}$ for simplicity.

\subsection{Associator of bimodule composition}
The associator $\lambda$ of bimodule composition $\circ$ is induced by the associator $\alpha$ of $\cB$ from the diagram below
\begin{equation}
\label{bimod_comp}
    \begin{tikzcd}
        (M\ot N) \ot P \arrow[r,"\alpha_{M,N,P}"]\ar[d,"{\ot[{A}](\ot[B]\,\ot \id_P)}"]
        &M\ot (N\ot P) \arrow[d,"{\ot[{B}](\id_M\ot \,\ot[A])}"]
        \\
        (M\circ N)\circ P=(M\ot[B]N)\ot[{A}]P \arrow[r,rightarrow,dashed, "\lambda_{M,N,P}"]
        &M\ot[{B}](N\ot[A]P)=M\circ(N\circ P)
    \end{tikzcd}
\end{equation}
It can be noticed that even the associator of $\cB$ is trivial, associator of bimodule composition is not necessarily trivial. This is because that the right $A$-action on $M \ot N$ and $M \circ N$ could be different, and so does the left $B$-action on $N \ot P$ and $N \circ P$, which may leads to a nontrivial $\lambda$.

\subsection{Associator bimodule and pentagonator}

For three objects $A,B,C\in \Sigma\cB$, the associator $\alpha_{A,B,C} :(A\ot B)\ot C\to A\ot (B\ot C) $ in $\cB$ is an algebra isomorphism. Therefore, in $\Sigma\cB$, we can define associator 1-morphisms as $\Lambda_{A,B,C}:={}_{A\ot (B\ot C)} (A\ot B)\ot C_{(A\ot B)\ot C}$, where the left module structure is induced by the algebra isomorphism $\alpha_{A,B,C}$. It is clear that $\Lambda_{A,B,C}$ is an invertible bimodule, and it is natural in $A,B,C$ following the naturality of $\alpha$. For example, for any bimodule $_D M_C$, the naturality of $\Lambda_{A, B, C}$ in $C$ leads to a 2-isomorphism
\begin{equation}
    \Lambda_{A,B,D}\circ ((A\ot B)\ot M) \stackrel{{\alpha}_{A,B,M}}{\longrightarrow} (A\ot (B\ot M)) \circ \Lambda_{A,B,C}.\label{eq.Lnatural}
\end{equation}

The pentagonator $\beta_{A,B,C,D}$ is a bimodule map between the following associator bimodule in $\Hom(((A\ot B)\ot C)\ot D, A\ot (B\ot (C\ot D)))$ in $\Sigma \cB$
\begin{equation}
    \begin{tikzcd}
    \Lambda_{A,B,C\ot D}\circ \Lambda_{A\ot B, C,D}\arrow[d,equal]
    & (A\ot \Lambda_{B,C,D}) \circ \Lambda_{A,B\ot C,D} \circ (\Lambda_{A,B,C}\ot D) \arrow[d,equal]
     \arrow[l,rightarrow,"\beta_{A,B,C,D}"'] 
    \\
     ((A\ot B)\ot C)\ot D \arrow[r,equal]
    & ((A\ot B)\ot C)\ot D
    \end{tikzcd}
\end{equation}
where the left module structure on the left hand side is induced by $\alpha_{A,B,C\ot D} \alpha_{A\ot B, C,D}$ and on the right hand side is induced by $(\id_A\ot \alpha_{B,C,D})  \alpha_{A,B\ot C,D}  (\alpha_{A,B,C}\ot \id_D)$. We omitted the associator  of bimodule composition here (in this example they are cancelled in the final result).  By the pentagon equation of $\cB$, the two bimodules are in fact equal to each other.
Therefore, the pentagonator $\beta_{A,B,C,D}$ is simply the identity bimodule map.

\subsection{Associator bimodule map}

As shown in eqn. \eqref{gen-asso-func}, the associator bimodule $\Lambda_{A,B,C}$ induce an equivalent functor
\begin{equation}
    \Lambda_{A,B,C}\circ -: \Hom(K,(A\ot B)\ot C) \to \Hom(K, A\ot (B\ot C)),
\end{equation}
which plays crucial roles in the calculation of 10j-symbols shown in eqn. \eqref{gen-def-10j} and will be studied in this subsection.

Since $\Sigma \cB$ is semisimple, we can focus on the case where all of $K, A, B, C$ are simple objects in $\Sigma \cB$.  Furthermore, the naturality of $\Lambda$ suggests that we only need to consider  representative objects chosen from each equivalent class of the simple objects.  Thus in the following, we consider only the objects in $\Sigma \cB_0$, a chosen set of representative objects in $\Sigma\cB$, and the bimodules in $\ho(A, (B \ot C) )$, a chosen set of representative simple $B\ot C$-$A$-bimodules for any $A,B,C\in \Sigma\cB_0$.

For any two separable algebras $A, B\in \Sigma\cB_0$, $A \ot B$ can be decomposed into direct sum of simple separable algebras in $\Sigma\cB_0$
\begin{align}
    A \ot B \cong \bigoplus_{M \in \Sigma\cB_0} F_M^{AB}  M,
\end{align}
where $F^{AB}_M:=\{s: M\to A\ot B, r: A\ot B\to M\}$ records the section and retraction algebra homomorphisms. We will drop $F$ for simplicity when it does not result in any confusion. It is clear that $A\ot B$ can be taken as an invertible $(A\ot B)$-$(\oplus M)$-bimodule, hence can be decomposed as
\begin{align}
A\ot B\cong \bigoplus_{M,Q} F^{AB}_{M;Q} {\ }_{A\ot B}Q_{M},
\end{align}
where $F^{AB}_{M;Q}:=\{s: Q\to A\ot B,r: A\ot B\to Q\}$ tracks section and retraction bimodule maps. Thus, any $(A\ot B) \ot C$-$K$-bimodule $U$ can be expressed as
\begin{align}
    U &\cong \bigoplus_{M, P, Q} (Q \otimes C) \circ P,
\end{align}
where $M\in\Sigma\cB_0$, $P\in \ho(K,M\ot C)$, $Q\in \ho(M, A\ot B)$, and the $F$s in the direct sum decomposition have been dropped for simplicity.  Therefore, we only need to study the $A\ot (B\ot C)$-$K$-bimodule $\Lambda_{A,B,C} \circ (Q \otimes C) \circ P$.

Similarly, any $A\ot (B\ot C)$-$K$-bimodule $V$ can be expressed as
\begin{align}
    V &\cong \bigoplus_{N, Y, X} (A \otimes Y) \circ X,
\end{align}
where $N\in \Sigma\cB_0$, $X\in \ho(K,A\ot N)$, $Y\in \ho(N,B\ot C)$.  Since $\Lambda_{A,B,C}\circ(Q\ot C)\circ P$ is a $A\ot (B\ot C)$-$K$-bimodule, it can be decomposed as
\begin{equation}
    \Lambda_{A,B,C}\circ(Q\ot C)\circ P \cong  \op[N,X,Y] F^{ABC; QP}_{KMN; YX} (A\ot Y)\circ X, \label{eq.sectionbasis}
\end{equation}
where $N\in \Sigma\cB_0$, $X\in \ho(K,A\ot N)$, $Y\in \ho(N,B\ot C)$. $F^{ABC; PQ}_{KMN; XY}$ tracks the section and retraction bimodule maps in the direct sum decomposition.
The normalized retraction bimodule maps serve as a basis for the calculation of the 10j-symbol (see Sec. \ref{sec:nor_sec_ret} for the normalization), while the corresponding normalized section bimodule maps are regarded as the dual basis, taken together they are referred to as associator bimodule maps.

\subsection{10j-symbol}
\label{sec:10j}

The 10j-symbol can be written down by fixing the choice of representative simple objects, simple 1-morphisms and bases of 2-morphisms (associator bimodule maps).
We consider the category $\Hom(K,A\ot (B\ot (C\ot D)))$ for any given $A,B,C,D,K\in \Sigma\cB_0$. The pentagonator induces a natural transformation between two equivalent functors, as depicted below
\begin{equation}
\label{pencirc}
    \begin{tikzcd}
        \Hom(K,((A\ot B)\ot C)\ot D) \arrow[ddd,bend right=50,"\Lambda_{A,B,C\ot D}\circ \Lambda_{A\ot B, C,D}\circ -"{left},""{name=L,right}] 
        \arrow[ddd,bend left=50,"(A\ot \Lambda_{B,C,D}) \circ \Lambda_{A,B\ot C,D} \circ (\Lambda_{A,B,C}\ot D)\circ -"{right},""{name=R,left}]
        \\ \\ \\ \Hom(K,A\ot (B\ot (C\ot D))) \arrow[rightarrow, from=R, to=L, "\beta_{A,B,C,D}\circ -"']
    \end{tikzcd}
\end{equation}
Although the pentagonator $\beta_{A,B,C,D}$ of $\Sigma\cB$ is trivial, the 10j-symbol, which characterizing the natural transformation induced by the pentagonator, is not necessarily trivial. This phenomenon is in analogy to that in group representation theory, the associator of $\Rep G$ is trivial but the 3j and 6j symbols are not trivial.

For any bimodule $U$ in $\Hom(K,((A\ot B)\ot C)\ot D)$, the natural transformation corresponds to a bimodule map between the image of the two functors, i.e. $\beta_{A, B, C, D} \circ U : (A\ot \Lambda_{B,C,D}) \circ \Lambda_{A,B\ot C,D} \circ (\Lambda_{A,B,C}\ot D)\circ U \rightarrow \Lambda_{A,B,C\ot D}\circ \Lambda_{A\ot B, C,D}\circ U$.  Since any bimodule in $\Hom(K,((A\ot B)\ot C)\ot D)$ can be decomposed as a direct sum of $((P_3\ot C)\ot D) \circ (P_2\ot D)\circ P_1$ for $M_1,M_2\in \Sigma\cB_0$ and $P_1\in \ho(K,M_1\ot D), P_2\in\ho(M_1,M_2\ot C), P_3\in \ho(M_2,A\ot B)$, we only need consider the case with $U = ((P_3\ot C)\ot D) \circ (P_2\ot D)\circ P_1$.

\begin{figure}[th]
\adjustbox{scale=0.7,center}{
\begin{tikzcd}
\itk{
    \coordinate (a)  at (-3,4);
    \coordinate (b)  at (-1,4);
    \coordinate (c)  at (1,4);
    \coordinate (d)  at (3,4);
    \coordinate (p) at (-2,3);
    \coordinate (x) at (-1,2);
    \coordinate (z)  at (0,1);
    \coordinate (y)  at (1,2);
    \coordinate (q)  at (2,3);
    \coordinate (w)  at (0,3);
    \coordinate (k)  at (0,0);
    \node[above] at (a) {$A$};
    \node[above] at (b) {$B$};
    \node[above] at (c) {$C$};
    \node[above] at (d) {$D$};
    \node[above] at (p) {$P_3$};
    \node[above] at (x) {$P_2$};
    \node[above] at (z) {$P_1$};
    \node[below] at (k) {$K$};
    \draw (a) -- (p) --node[below left]{$M_2$} (x) --node[below left]{$M_1$} (z) -- (k);
    \draw (b) -- (p);
    \draw (c) -- (x);
    \draw (d) -- (z);
}
\arrow[rightarrow,d,"","\mid \tilde{\zeta}^1\rangle"']
\arrow[rightarrow,ddd,bend right=40,"", "\mid Z \rangle"'{name=U}]
&&
\itk{
    \coordinate (a)  at (-3,4);
    \coordinate (b)  at (-1,4);
    \coordinate (c)  at (1,4);
    \coordinate (d)  at (3,4);
    \coordinate (p) at (-2,3);
    \coordinate (x) at (-1,2);
    \coordinate (z)  at (0,1);
    \coordinate (y)  at (1,2);
    \coordinate (q)  at (2,3);
    \coordinate (w)  at (0,3);
    \coordinate (k)  at (0,0);
    \node[above] at (a) {$A$};
    \node[above] at (b) {$B$};
    \node[above] at (c) {$C$};
    \node[above] at (d) {$D$};
    \node[above] at (p) {$P_3$};
    \node[above] at (x) {$P_2$};
    \node[above] at (z) {$P_1$};
    \node[below] at (k) {$K$};
    \draw (a) -- (p) --node[below left]{$M_2$} (x) --node[below left]{$M_1$} (z) -- (k);
    \draw (b) -- (p);
    \draw (c) -- (x);
    \draw (d) -- (z);
}
\arrow[d,rightarrow,"\mid\tilde{\zeta}^3\rangle"]
\arrow[rightarrow,ddd,bend left=40,"\mid YWXJ\rangle"{name=N}]
\arrow[rightarrow,ll,"\beta"]
\\
\itk{
    \coordinate (a)  at (-3,4);
    \coordinate (b)  at (-1,4);
    \coordinate (c)  at (1,4);
    \coordinate (d)  at (3,4);
    \coordinate (p) at (-2,3);
    \coordinate (x) at (-1,2);
    \coordinate (z)  at (0,1);
    \coordinate (y)  at (1,2);
    \coordinate (q)  at (2,3);
    \coordinate (w)  at (0,3);
    \coordinate (k)  at (0,0);
    \node[above] at (a) {$A$};
    \node[above] at (b) {$B$};
    \node[above] at (c) {$C$};
    \node[above] at (d) {$D$};
    \node[above] at (p) {$P_3$};
    \node[above left] at (y) {$Q_3$};
    \node[above] at (z) {$Z$};
    \node[below] at (k) {$K$};
    \draw (a) -- (p) --node[left]{$M_2$} (z) -- (k);
    \draw (b) -- (p);
    \draw (c) -- (y);
    \draw (d) -- (y) --node[below right]{$N_2$} (z);
} \arrow[rightarrow,d,"","\text{Ic}"']
&&
\itk{
    \coordinate (a)  at (-3,4);
    \coordinate (b)  at (-1,4);
    \coordinate (c)  at (1,4);
    \coordinate (d)  at (3,4);
    \coordinate (p) at (-2,3);
    \coordinate (x) at (-1,2);
    \coordinate (z)  at (0,1);
    \coordinate (y)  at (1,2);
    \coordinate (q)  at (2,3);
    \coordinate (w)  at (0,3);
    \coordinate (k)  at (0,0);
    \node[above] at (a) {$A$};
    \node[above] at (b) {$B$};
    \node[above] at (c) {$C$};
    \node[above] at (d) {$D$};
    \node[above] at (w) {$W$};
    \node[above] at (x) {$X$};
    \node[above] at (z) {$P_1$};
    \node[below] at (k) {$K$};
    \draw (a) -- (x)  --node[below left]{$M_1$} (z) -- (k);
    \draw (b) -- (w);
    \draw (c) -- (w) --node[right]{$J$} (x);
    \draw (d) -- (z);
}
\arrow[d,rightarrow,"\mid \tilde{\zeta}^4\rangle"]
\\
\itk{
    \coordinate (a)  at (-3,4);
    \coordinate (b)  at (-1,4);
    \coordinate (c)  at (1,4);
    \coordinate (d)  at (3,4);
    \coordinate (p) at (-2,3);
    \coordinate (x) at (-1,2);
    \coordinate (z)  at (0,1);
    \coordinate (y)  at (1,2);
    \coordinate (q)  at (2,3);
    \coordinate (w)  at (0,3);
    \coordinate (k)  at (0,0);
    \node[above] at (a) {$A$};
    \node[above] at (b) {$B$};
    \node[above] at (c) {$C$};
    \node[above] at (d) {$D$};
    \node[above right] at (x) {$P_3$};
    \node[above] at (q) {$Q_3$};
    \node[above] at (z) {$Z$};
    \node[below] at (k) {$K$};
    \draw (a) -- (x) --node[below left]{$M_2$} (z) -- (k);
    \draw (b) -- (x);
    \draw (c) -- (q);
    \draw (d) -- (q) --node[right]{$N_2$} (z);
} 
\arrow[rightarrow,d,"","\mid \tilde{\zeta}^2 \rangle"']
&&
\itk{
    \coordinate (a)  at (-3,4);
    \coordinate (b)  at (-1,4);
    \coordinate (c)  at (1,4);
    \coordinate (d)  at (3,4);
    \coordinate (p) at (-2,3);
    \coordinate (x) at (-1,2);
    \coordinate (z)  at (0,1);
    \coordinate (y)  at (1,2);
    \coordinate (q)  at (2,3);
    \coordinate (w)  at (0,3);
    \coordinate (k)  at (0,0);
    \node[above] at (a) {$A$};
    \node[above] at (b) {$B$};
    \node[above] at (c) {$C$};
    \node[above] at (d) {$D$};
    \node[above] at (w) {$W$};
    \node[above] at (y) {$Y$};
    \node[above] at (z) {$Q_1$};
    \node[below] at (k) {$K$};
    \draw (a) -- (z) -- (k);
    \draw (b) -- (w);
    \draw (c) -- (w) --node[left]{$J$} (y);
    \draw (d) -- (y) --node[below right]{$N_1$} (z);
}
\arrow[d,rightarrow,"\mid \tilde{\zeta}^5 \rangle"]
\\ 
\itk{
    \coordinate (a)  at (-3,4);
    \coordinate (b)  at (-1,4);
    \coordinate (c)  at (1,4);
    \coordinate (d)  at (3,4);
    \coordinate (p) at (-2,3);
    \coordinate (x) at (-1,2);
    \coordinate (z)  at (0,1);
    \coordinate (y)  at (1,2);
    \coordinate (q)  at (2,3);
    \coordinate (w)  at (0,3);
    \coordinate (k)  at (0,0);
    \node[above] at (a) {$A$};
    \node[above] at (b) {$B$};
    \node[above] at (c) {$C$};
    \node[above] at (d) {$D$};
    \node[above] at (y) {$Q_2$};
    \node[above] at (q) {$Q_3$};
    \node[above] at (z) {$Q_1$};
    \node[below] at (k) {$K$};
    \draw (a) -- (z) -- (k);
    \draw (b) -- (y);
    \draw (c) -- (q);
    \draw (d) -- (q) --node[below right]{$N_2$} (y) --node[below right]{$N_1$} (z);
} 
\arrow[rightarrow,rr,"g"]
&&
\itk{
    \coordinate (a)  at (-3,4);
    \coordinate (b)  at (-1,4);
    \coordinate (c)  at (1,4);
    \coordinate (d)  at (3,4);
    \coordinate (p) at (-2,3);
    \coordinate (x) at (-1,2);
    \coordinate (z)  at (0,1);
    \coordinate (y)  at (1,2);
    \coordinate (q)  at (2,3);
    \coordinate (w)  at (0,3);
    \coordinate (k)  at (0,0);
    \node[above] at (a) {$A$};
    \node[above] at (b) {$B$};
    \node[above] at (c) {$C$};
    \node[above] at (d) {$D$};
    \node[above] at (y) {$Q_2$};
    \node[above] at (q) {$Q_3$};
    \node[above] at (z) {$Q_1$};
    \node[below] at (k) {$K$};
    \draw (a) -- (z) -- (k);
    \draw (b) -- (y);
    \draw (c) -- (q);
    \draw (d) -- (q) --node[below right]{$N_2$} (y) --node[below right]{$N_1$} (z);
} 
\end{tikzcd}} 
\caption{$\mid Z \rangle$ and $\mid YWXJ \rangle$ are two different retraction bimodule maps.}
\label{eq.10j}
\end{figure}
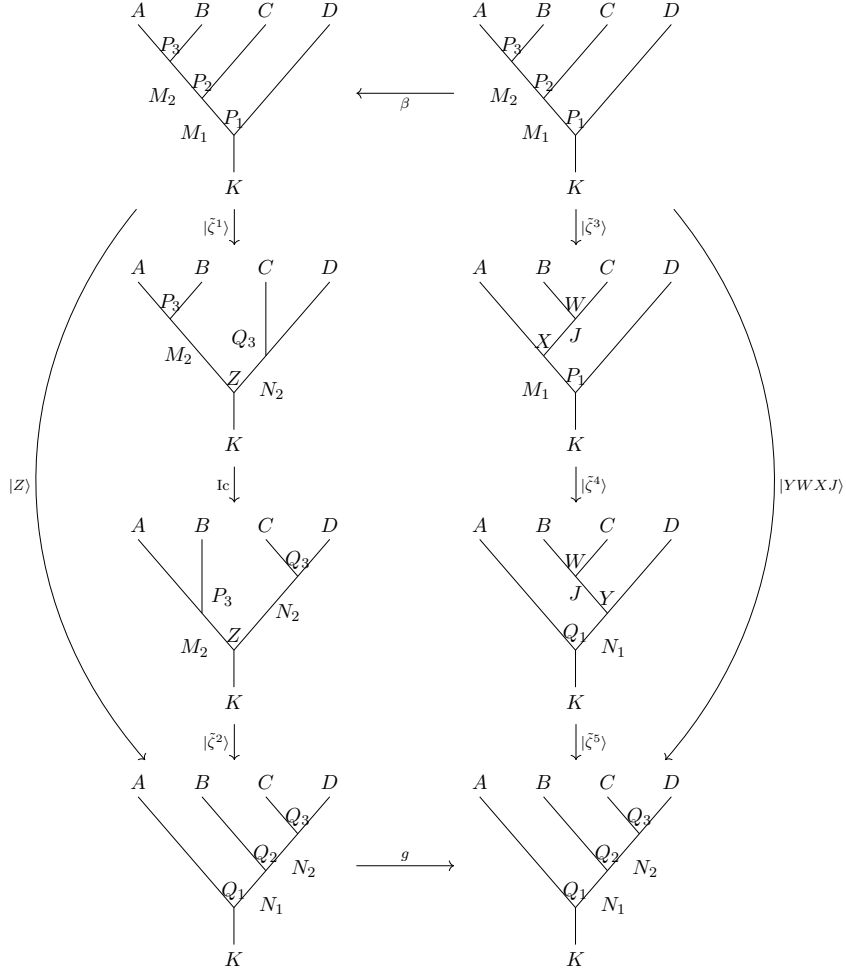

We denote $V_1 \equiv \Lambda_{A,B,C\ot D}\circ \Lambda_{A\ot B, C,D}\circ U$ and $V_2 \equiv (A\ot \Lambda_{B,C,D}) \circ \Lambda_{A,B\ot C,D} \circ (\Lambda_{A,B,C}\ot D)\circ U$, and they are objects in $\Hom(K,A\ot (B\ot (C\ot D)))$.  Any $A\ot (B\ot (C\ot D))$-$K$-bimodule can be expressed as direct sum of $(A\ot (B\ot Q_3)) \ci (A\ot Q_2)\ci Q_1$, with $N_1,N_2\in \Sigma\cB_0$, $Q_1\in \ho(K,A \ot N_1), Q_2\in\ho(N_1,B\ot N_2), Q_3\in \ho(N_2,C\ot D)$. Thus, the bimodule map $\beta_{A,B,C,D} \ci U$ reduces to an endomorphism $g^{A,B,C,D,U}_{N_1,N_2;Q_1,Q_2,Q_3}$ of bimodule $A\ot (B\ot Q_3)) \ci (A\ot Q_2)\ci Q_1 $ satisfies $g \cdot \cl {Z} \cdot \beta = \cl {YWXJ}$, where $\cl {Z}$ and $\cl {YWXJ}$ are normalized retraction maps (see below for details) in the direct sum decomposition of $V_1$ and $V_2$, respectively. Since the pentagonator $\beta_{A, B, C, D}$ is trivial, i.e. $\beta = \id$, we have $V_1 = V_2 = V$ and
\begin{align}
\label{eq:g}
    \cl {YWXJ} = g \cdot \cl {Z}.
\end{align}
Therefore, the 10j-symbols, which are characterized by $g$, are determined by the two direct sum compositions of $V$, where the first decomposition is given below
\begin{equation}
\begin{tikzcd}
    \Lambda_{A,B,C\ot D}\circ \Lambda_{A\ot B, C,D}\circ  ((P_3\ot C)\ot D) \ci[1] (P_2\ot D)\ci[2] P_1
    \arrow[d,rightarrow,"\lambda"]
    \arrow[rightarrow, "\oplus \mid \tilde{\zeta}^1 \rangle"',bend right,shift right=31ex,ddd]
    \\ \Lambda_{A,B,C\ot D}\circ \Lambda_{A\ot B, C,D}\circ  ((P_3\ot C)\ot D) \ci[2] (P_2\ot D)\ci[1] P_1
    \arrow[d,rightarrow,"\alpha_{P_3,C,D}"]
    \\ \Lambda_{A,B,C\ot D}\circ (P_3\ot (C\ot D)) \ci[2] \Lambda_{M_2,C,D} \circ (P_2\ot D)\ci[1] P_1
    \arrow[d,rightarrow, "\oplus \mid {\zeta}^1 \rangle"]
    \\ \op[N_2, Z, Q_3] F^{M_2CD;P_2P_1}_{KM_1N_2;Q_3Z} \Lambda_{A,B,C\ot D}\circ (P_3\ot (C\ot D)) \ci[2]  (M_2\ot Q_3)\ci[1] Z
    \arrow[rightarrow,"\text{Ic}"',bend right,shift right=31ex,dd]
    \arrow[d,rightarrow,"\lambda"]
    \\ \op[N_2, Z, Q_3] F^{M_2CD;P_2P_1}_{KM_1N_2;Q_3Z} \Lambda_{A,B,C\ot D}\circ (P_3\ot (C\ot D)) \ci[1]  (M_2\ot Q_3)\ci[2] Z
    \arrow[d,rightarrow,"\phi_{P_3, Q_3}"]
    \\ \op[N_2, Z, Q_3]F^{M_2CD;P_2P_1}_{KM_1N_2;Q_3Z} \Lambda_{A,B,C\ot D}\circ ((A\ot B)\ot Q_3)\ci[1] (P_3\ot N_2) \ci[2] Z
    \arrow[d,rightarrow,"\lambda"]
    \arrow[rightarrow, "\oplus \mid \tilde{\zeta}^2 \rangle"',bend right,shift right=31ex,ddd]
    \\ \op[N_2, Z, Q_3]F^{M_2CD;P_2P_1}_{KM_1N_2;Q_3Z} \Lambda_{A,B,C\ot D}\circ ((A\ot B)\ot Q_3)\ci[2] (P_3\ot N_2) \ci[1] Z
    \arrow[d,rightarrow,"\alpha_{A,B,Q_3}"]
    \\ \op[N_2, Z, Q_3]F^{M_2CD;P_2P_1}_{KM_1N_2;Q_3Z} (A\ot (B\ot Q_3)) \ci[2] \Lambda_{A,B,N_2}\circ (P_3\ot N_2)  \ci[1] Z
    \arrow[d,rightarrow, "\oplus \mid {\zeta}^2 \rangle"]
    \\ \op[N_1,N_2, Q_1,Q_2, Q_3] \co[Z] F^{M_2CD;P_2P_1}_{KM_1N_2;Q_3Z}F^{ABN_2;P_3Z}_{KM_2N_1;Q_2Q_1} (A\ot (B\ot Q_3)) \ci[2] (A\ot Q_2)\ci[1] Q_1.
\end{tikzcd}
\label{eq.sd}
\end{equation}
We have used the naturality of $\Lambda$ \eqref{eq.Lnatural}, the interchanger \eqref{eq.interchange}, and the decomposition \eqref{eq.sectionbasis}.  $\ci[1]$ means the composition should be done firstly and $\ci[2]$ means the composition should be done secondly. $\lambda$ is the associator of the composition of three bimodules defined in eqn. \eqref{bimod_comp}. $\alpha$ is the associator of the tensor products of three bimodules.  $F^{M_2CD;P_2P_1}_{KM_1N_2;Q_3Z}$, and $F^{ABN_2;P_3Z}_{KM_2N_1;Q_2Q_1}$ tracks the corresponding direct sum decompositions, while $\mid {\zeta^1} \rangle$ and $\mid {\zeta^2} \rangle$ are the normalized retractions defined in Sec. \ref{sec:nor_sec_ret} (the corresponding normalized sections are denoted as $\langle {\zeta^1} \mid$ and $\langle {\zeta^2} \mid$, respectively).  
Note that we leave the identity maps implicit and only write the vital step in the equation.
For simplicity, we introduce two maps $\mid \tilde{\zeta}^1 \rangle$ and $\mid \tilde{\zeta}^2 \rangle$ as shown in the equation, and hence the above decomposition can be depicted as left path in fig.\ref{eq.10j}.

Similarly, the second direct sum decomposition is given by
\begin{equation}
\begin{tikzcd}
    (A\ot \Lambda_{B,C,D}) \circ \Lambda_{A,B\ot C,D} \circ (\Lambda_{A,B,C}\ot D)\circ((P_3\ot C)\ot D) \ci[1] (P_2\ot D)\ci[2] P_1
    \arrow[d,rightarrow,"\tilde{c}_{D,P_2}"]
    \arrow[rightarrow,ddddd,"\oplus \mid \tilde{\zeta}^3 \rangle",bend right,shift right=40ex]
    \\ (A\ot \Lambda_{B,C,D}) \circ \Lambda_{A,B\ot C,D} \circ (\Lambda_{A,B,C}\ot D)\circ(((P_3\ot C)\ci[1] P_2) \ot (D\circ D))\ci[2] P_1
    \arrow[d,rightarrow,"\tilde{c}_{D,(P_3\ot C)\circ P_2}"]
    \\ (A\ot \Lambda_{B,C,D}) \circ \Lambda_{A,B\ot C,D} \circ ((\Lambda_{A,B,C}\circ(P_3\ot C)\ci[1] P_2) \ot D)\ci[2] P_1
    \arrow[rightarrow,d, "\oplus \mid {\zeta^3} \rangle"]
    \\ \op[J,X,W]F^{ABC;P_3P_2}_{M_1M_2J;WX} (A\ot \Lambda_{B,C,D}) \circ \Lambda_{A,B\ot C,D} \circ (((A\ot W)\ci[1] X) \ot D)\ci[2] P_1
    \arrow[d,rightarrow,"\tl{c}^{-1}_{D,X}"]
    \\ \op[J,X,W]F^{ABC;P_3P_2}_{M_1M_2J;WX} (A\ot \Lambda_{B,C,D}) \circ \Lambda_{A,B\ot C,D} \circ ((A\ot W)\ot D) \ci[1](X\ot D)\ci[2] P_1
    \arrow[d,rightarrow,"\lambda"]
    \\ \op[J,X,W]F^{ABC;P_3P_2}_{M_1M_2J;WX} (A\ot \Lambda_{B,C,D}) \circ \Lambda_{A,B\ot C,D} \circ ((A\ot W)\ot D) \ci[2](X\ot D)\ci[1] P_1
    \arrow[d,rightarrow,"\alpha_{A,W,D}"]
    \arrow[rightarrow, "\oplus \mid \tilde{\zeta}^4 \rangle",bend right,shift right=40ex,ddd]
    \\ \op[J,X,W]F^{ABC;P_3P_2}_{M_1M_2J;WX} (A\ot \Lambda_{B,C,D}) \circ (A\ot (W\ot D))\ci[2] \Lambda_{A,J,D}  \circ(X\ot D)\ci[1] P_1
    \arrow[rightarrow,d, "\oplus \mid {\zeta}^4 \rangle"]
    \\ \op[J,W,N_1,Q_1,Y]\co[X] F^{ABC;P_3P_2}_{M_1M_2J;WX}F^{AJD;XP_1}_{KM_1N_1;YQ_1} (A\ot \Lambda_{B,C,D}) \circ (A\ot (W\ot D))\ci[2] (A\ot Y) \ci[1] Q_1
    \arrow[d,rightarrow,"\lambda"]
    \\ \op[J,W,N_1,Q_1,Y]\co[X] F^{ABC;P_3P_2}_{M_1M_2J;WX}F^{AJD;XP_1}_{KM_1N_1;YQ_1} (A\ot \Lambda_{B,C,D}) \circ (A\ot (W\ot D))\ci[1] (A\ot Y) \ci[2] Q_1
    \arrow[d,rightarrow,"\tl{c}_{\Lambda_{B,C,D},A}"]
    \arrow[rightarrow, "\oplus \mid \tilde{\zeta}^5\rangle",bend right,shift right=40ex,ddddd]
    \\ \op[J,W,N_1,Q_1,Y]\co[X] F^{ABC;P_3P_2}_{M_1M_2J;WX}F^{AJD;XP_1}_{KM_1N_1;YQ_1} (A\ot (\Lambda_{B,C,D} \circ (W\ot D))\ci[1] (A\ot Y)\ci[2] Q_1
    \arrow[d,rightarrow,"\tl{c}_{\Lambda_{B,C,D} \circ (W\ot D),A}"]
    \\ \op[J,W,N_1,Q_1,Y]\co[X] F^{ABC;P_3P_2}_{M_1M_2J;WX}F^{AJD;XP_1}_{KM_1N_1;YQ_1} (A\ot (\Lambda_{B,C,D} \circ (W\ot D)\ci[1] Y))\ci[2] Q_1
    \arrow[rightarrow,d, "\oplus \mid {\zeta^5} \rangle"]
    \\ \op[N_1,N_2,Q_1,Q_2,Q_3]\co[J,W,X,Y] F^{ABC;P_3P_2}_{M_1M_2J;WX}F^{AJD;XP_1}_{KM_1N_1;YQ_1}F^{BCD;WY}_{N_1JN_2;Q_3Q_2} (A\ot ((B\ot Q_3)\ci[1] Q_2)\ci[2] Q_1
    \arrow[d,rightarrow,"\tl{c}^{-1}_{B\ot Q_3,A}"]
    \\ \op[N_1,N_2,Q_1,Q_2,Q_3]\co[J,W,X,Y] F^{ABC;P_3P_2}_{M_1M_2J;WX}F^{AJD;XP_1}_{KM_1N_1;YQ_1}F^{BCD;WY}_{N_1JN_2;Q_3Q_2} (A\ot (B\ot Q_3))\ci[1] (A\ot Q_2)\ci[2] Q_1
    \arrow[d,rightarrow,"\lambda"]
    \\ \op[N_1,N_2,Q_1,Q_2,Q_3]\co[J,W,X,Y] F^{ABC;P_3P_2}_{M_1M_2J;WX}F^{AJD;XP_1}_{KM_1N_1;YQ_1}F^{BCD;WY}_{N_1JN_2;Q_3Q_2} (A\ot (B\ot Q_3))\ci[2] (A\ot Q_2)\ci[1] Q_1.
\end{tikzcd}
\label{eq.ld}
\end{equation}
In this decomposition we used interchange law several times, which will cancel each other in the end due to the naturality of braiding $c$.  The decomposition can be depicted as the right path in the fig. \ref{eq.10j} with the maps $\mid \tilde{\zeta}^3\rangle$, $\mid\tilde{\zeta}^4\rangle$, and $\mid \tilde{\zeta}^5\rangle$.

Furthermore, we introduce two bimodule maps $\cl Z$ and $\cl {YWXJ}$
\begin{align}
\cl Z: & V_1 \rightarrow A\ot (B\ot Q_3)) \circ (A\ot Q_2)\circ Q_1, \nonumber\\
\cl {YWXJ}:& V_2 \rightarrow  (A\ot (B\ot Q_3)) \circ (A\ot Q_2)\circ Q_1.
\end{align}
as shown in the figure.  $\cl Z$ is the composition of the three bimodule maps, $\cl {\tilde{\zeta}^1}$, $\mathrm{Ic}$, and $\cl {\tilde{\zeta}^2}$, which is determined by the 1-morphism $Z$ for given $P_i$ and $Q_i$.
Here we consider only those ${\zeta^1}$ and ${\zeta^2}$ that is valid in fig. \ref{eq.10j}, i.e. they share the same 1-morphism $Z$. As a result, ${\zeta^1}$ and ${\zeta^2}$ is also uniquely determined by $Z$.
$\cl {YWXJ}$ is the composition of $\cl {\tilde{\zeta}^3}$, $\cl {\tilde{\zeta}^4}$, $\cl {\tilde{\zeta}^5}$ and fully determined by the 1-morphisms $W$, $X$, $Y$ and object $J$. Similarly, we consider only those ${\zeta^3}$, ${\zeta^4}$ and ${\zeta^5}$ sharing $Y$, $W$, $X$ and $J$. Therefore, ${\zeta^3}$, ${\zeta^4}$ and ${\zeta^5}$ is uniquely determined by $Y$, $W$, $X$ and $J$.
$\cl Z$ and $\cl {YWXJ}$ can be regarded as two different bases of the vector space of the bimodule maps from $V$ to $(A\ot (B\ot Q_3)) \circ (A\ot Q_2)\circ Q_1$.

According to eqn. \eqref{eq:g} and fig. \ref{eq.10j}, the basis transformation can be expressed as
\begin{align}
\label{eq:10j-G}
   \mid YWXJ  \rangle = g \cdot \mid Z \rangle = \sum_{Z} G_{Z}^{YWXJ} \mid Z \rangle.
\end{align}
where $G_{Z}^{YWXJ}$ is just the 10j-symbol.  Since $\mid YWXJ \rangle$ is over-complete, the above transformation and  the 10j-symbol $G_{Z}^{YWXJ}$ as a matrix is non-invertible. However, we can define its right inverse as
\begin{align}
\label{eq:10j-G--1}
    \mid {Z} \rangle = \sum_{YWXJ} {(G^{-1})^{Z}_{YWXJ}} \mid {YWXJ} \rangle,
\end{align}
where $\sum_{YWXJ} G_{Z}^{YWXJ} (G^{-1})_{YWXJ}^{Z'} = \delta_{Z}^{Z'}$.

In practice, $\mid {\zeta^1} \rangle$, $\mid {\zeta^2} \rangle$, $\mid {\zeta^3} \rangle$, $\mid {\zeta^4} \rangle$ and $\mid {\zeta^5} \rangle$ are the basis of the vector spaces assigned to the five 3-simplices of the boundary of a 4-simplex, and the 10j-symbol $G_{Z}^{YWXJ}$ is the data assigned to the 4-simplex.

\section{Spherical Structure of \texorpdfstring{$\Sigma\cB$}{ΣB}}
\label{sVecspher}
In this section, we will introduce the spherical structure of $\Sigma\cB$, which plays a crucial role in our construction.  With the spherical structure, we can define a pairing $[ \rho, \xi ]$ between the bimodule maps $\xi: f \rightarrow g$ and $\rho: g \rightarrow f$ with $f, g \in \hom(A, B)$, which is very useful in calculating the 10j-symbols.  We can also define and calculate the quantum dimensions of objects and 1-morphisms, which in together determine the normalization factor of the 10j-symbols. Instead of providing a strict definition, we only introduce some properties of the spherical structure which are related to our paper. Please check Ref.~\cite{douglas2018} for a detailed definition.

\subsection{Spherical structure}

In a spherical fusion 2-category, every object has a left and a right dual, every 1-morphism has a left and a right adjoint and every 2-morphism has a left and a right mate.
For an object $A$, its right dual is a triple $(A^\star, \ e_A: A\ot A^\star \rightarrow \1, \ i_A: \1 \rightarrow A^\star \ot A)$ where $A^\star$ is an object and $e_B$, $i_B$ are two 1-morphisms (called folds). 
For a 1-morphism $f:A\to B$, its right adjoint is a triple $(f^*:B\to A, \ \eta_f: \id_A  \rightarrow f^\ast \circ f, \ \epsilon_f: f^\ast \circ f \rightarrow \id_B)$. $f^\ast$ is a 1-morphism and $\eta_f,\epsilon_f$ are two 2-morphisms satisfying the cusp equations
\begin{align}
    &(\epsilon_f \circ \id_f) \cdot (\id_f \circ \eta_f)=\id_f, &
    &(\id_{f^\ast}  \circ \epsilon_f) \cdot (\eta_f \circ \id_{f^\ast} ) =\id_{f^\ast},
\end{align}
which can be graphically expressed as
\begin{align*}
\label{graph of cusp}
\begin{array}{c}
\begin{tikzpicture}[scale=3]
\coordinate[label=right:$ $] (0') at (2,0) ;
\coordinate[label=above:$ $] (1') at (2,1) ;
\coordinate[label=left:$ $] (2')at (3,0) ;
\coordinate[label=below:$ $] (3') at (3,1) ;
\draw[help lines] (0')--(1') node[midway,right] {$B$};%
\draw[very thick] (2.5,1)--(2.5,0.5);
\draw[->,very thick] (2.5,0)--(2.5,0.5) node[midway,left] {$f$};
\draw[help lines] (0')--(2') node[midway,left] {$ $};%
\draw[help lines] (3')--(1') node[midway,left] {$ $};%
\draw[help lines] (3')--(2') node[midway,left] {$A$};%
\end{tikzpicture}
\end{array}
=
\begin{array}{c}
\begin{tikzpicture}[scale=3]
\coordinate[label=right:$ $] (0') at (2,0) ;
\coordinate[label=above:$ $] (1') at (2,1) ;
\coordinate[label=left:$ $] (2')at (3,0) ;
\coordinate[label=below:$ $] (3') at (3,1) ;
\draw[help lines] (0')--(1') node[midway,right] {$B$};%
\draw[->,very thick] (2.25,0)--(2.25,0.5) node[midway,left] {$f$};
\draw[->,very thick] (2.25,0.5) arc(180:0:0.125);
\draw[->,very thick] (2.5,0.5) arc(180:360:0.125);
\draw[very thick] (2.75,0.5)--(2.75,1) node[midway,right] {$f$};
\draw[dashed] (2.625,0)--(2.625,0.375) node[right=4pt] {$\eta_f$};;
\draw[dashed] (2.375,1)--(2.375,0.625) node[left=4pt] {$\epsilon_f$};
\draw[help lines] (0')--(2') node[midway,left] {$ $};%
\draw[help lines] (3')--(1') node[midway,left] {$ $};%
\draw[help lines] (3')--(2') node[midway,left] {$A$};%
\end{tikzpicture}
\end{array}
,\quad
\begin{array}{c}
\begin{tikzpicture}[scale=3]
\coordinate[label=right:$ $] (0') at (2,0) ;
\coordinate[label=above:$ $] (1') at (2,1) ;
\coordinate[label=left:$ $] (2')at (3,0) ;
\coordinate[label=below:$ $] (3') at (3,1) ;
\draw[help lines] (0')--(1') node[midway,right] {$A$};%
\draw[->,very thick] (2.5,1)--(2.5,0.5);
\draw[very thick] (2.5,0)--(2.5,0.5) node[midway,left] {$f\ast$};
\draw[help lines] (0')--(2') node[midway,left] {$ $};%
\draw[help lines] (3')--(1') node[midway,left] {$ $};%
\draw[help lines] (3')--(2') node[midway,left] {$B$};%
\end{tikzpicture}
\end{array}
=
\begin{array}{c}
\begin{tikzpicture}[scale=3]
\coordinate[label=right:$ $] (0') at (2,0) ;
\coordinate[label=above:$ $] (1') at (2,1) ;
\coordinate[label=left:$ $] (2')at (3,0) ;
\coordinate[label=below:$ $] (3') at (3,1) ;
\draw[help lines] (0')--(1') node[midway,right] {$A$};%
\draw[->,very thick] (2.25,1)--(2.25,0.5) node[midway,left] {$f^\ast$};
\draw[->,very thick] (2.25,0.5) arc(180:360:0.125);
\draw[->,very thick] (2.5,0.5) arc(180:0:0.125);
\draw[very thick] (2.75,0.5)--(2.75,0) node[midway,right] {$f^\ast$};
\draw[dashed] (2.375,0)--(2.375,0.375) node[left=4pt] {$\eta_f$};;
\draw[dashed] (2.625,1)--(2.625,0.625) node[right=4pt] {$\epsilon_f$};
\draw[help lines] (0')--(2') node[midway,left] {$ $};%
\draw[help lines] (3')--(1') node[midway,left] {$ $};%
\draw[help lines] (3')--(2') node[midway,left] {$B$};%
\end{tikzpicture}
\end{array}
\end{align*}
All the duals and adjoints are involutive, i.e. $f=f^{\ast \ast}$, $B=B^{\star \star}$. For a 2-morphism $\sigma: f \rightarrow g$, its left/right mates agree, i.e.
\begin{align}
\sigma=(\epsilon_f \circ \id_g) \cdot (\id_f \circ \sigma^\ast \circ \id_g) \cdot (\id_f \circ \eta_g)= (\id_g \circ \epsilon_{f^\ast}) \cdot (\id_g \circ \sigma^\ast \circ \id_f) \cdot (\eta_{g^\ast} \circ \id_f)
\end{align}
which can be graphically expressed as
\begin{align*}
\begin{array}{c}
\begin{tikzpicture}[scale=3]
\coordinate[label=right:$ $] (0') at (2,0) ;
\coordinate[label=above:$ $] (1') at (2,1) ;
\coordinate[label=left:$ $] (2')at (3,0) ;
\coordinate[label=below:$ $] (3') at (3,1) ;
\draw[help lines] (0')--(1') node[midway,right] {$B$};%
\draw[very thick] (2.5,0.7)--(2.5,1) node[midway,left] {$g$};
\draw[->,very thick] (2.5,0)--(2.5,0.3) node[midway,left] {$f$};
\fill (2.5,0.5) circle (0.6pt) node[left] {$\sigma$};
\draw[->,very thick] (2.5,0.3)--(2.5,0.7);
\draw[help lines] (0')--(2') node[midway,left] {$ $};%
\draw[help lines] (3')--(1') node[midway,left] {$ $};%
\draw[help lines] (3')--(2') node[midway,left] {$A$};%
\end{tikzpicture}
\end{array}
=
\begin{array}{c}
\begin{tikzpicture}[scale=3]
\coordinate[label=right:$ $] (0') at (2,0) ;
\coordinate[label=above:$ $] (1') at (2,1) ;
\coordinate[label=left:$ $] (2')at (3,0) ;
\coordinate[label=below:$ $] (3') at (3,1) ;
\draw[help lines] (0')--(1') node[midway,right] {$B$};%
\draw[->,very thick] (2.25,0)--(2.25,0.5) node[midway,left] {$f$};
\draw[very thick] (2.25,0.5) arc(180:0:0.125);
\fill (2.5,0.5) circle (0.6pt) node[left] {$\sigma^\ast$};
\draw[->,very thick] (2.5,0.5) arc(180:360:0.125);
\draw[very thick] (2.75,0.5)--(2.75,1) node[midway,right] {$g$};
\draw[dashed] (2.625,0)--(2.625,0.375) node[right=4pt] {$\eta_g$};
\draw[dashed] (2.375,1)--(2.375,0.625) node[left=4pt] {$\epsilon_f$};
\draw[help lines] (0')--(2') node[midway,left] {$ $};%
\draw[help lines] (3')--(1') node[midway,left] {$ $};%
\draw[help lines] (3')--(2') node[midway,left] {$A$};%
\end{tikzpicture}
\end{array}
=
\begin{array}{c}
\begin{tikzpicture}[scale=3]
\coordinate[label=right:$ $] (0') at (2,0) ;
\coordinate[label=above:$ $] (1') at (2,1) ;
\coordinate[label=left:$ $] (2')at (3,0) ;
\coordinate[label=below:$ $] (3') at (3,1) ;
\draw[help lines] (0')--(1') node[midway,right] {$B$};%
\draw[very thick] (2.25,1)--(2.25,0.5) node[midway,left] {$g$};
\fill (2.5,0.5) circle (0.6pt) node[left] {$\sigma^\ast$};
\draw[->,very thick] (2.5,0.5) arc(360:180:0.125);
\draw[very thick] (2.75,0.5) arc(0:180:0.125);
\draw[->,very thick] (2.75,0)--(2.75,0.5) node[midway,right] {$f$};
\draw[dashed] (2.375,0)--(2.375,0.375) node[left=4pt] {$\eta_{g^\ast}$};
\draw[dashed] (2.625,1)--(2.625,0.625) node[right=4pt]{$\epsilon_{f^\ast}$};
\draw[help lines] (0')--(2') node[midway,left] {$ $};%
\draw[help lines] (3')--(1') node[midway,left] {$ $};%
\draw[help lines] (3')--(2') node[midway,left] {$A$};%
\end{tikzpicture}
\end{array}
\end{align*}

Then one can define the left planar trace $\mathrm{Tr_L}(\xi): \id_A \rightarrow \id_A$ and right planar trace $\mathrm{Tr_R}(\xi): \id_B \rightarrow \id_B$ of any 2-endomorphism $\xi: f \rightarrow f $ for any arbitrary 1-morphism $f: A \rightarrow B$ as
\begin{align}
&\mathrm{Tr_L}(\xi):=\epsilon_{f^\ast} \cdot (\id_{f^\ast} \circ \xi ) \cdot \eta_f=
\begin{array}{c}
\begin{tikzpicture}[scale=3]
\coordinate[label=above:$B$] (ct) at (2.5,0.4);
\coordinate[label=above:$A$] (cn) at (2.9,0);
\coordinate[label=right:$ $] (0') at (2,0) ;
\coordinate[label=above:$ $] (1') at (2,1) ;
\coordinate[label=left:$ $] (2')at (3,0) ;
\coordinate[label=below:$ $] (3') at (3,1) ;
\draw[help lines] (0')--(1');
\draw[dashed] (2.5,0)--(2.5,0.2) node[midway,left=2pt] {$\eta_f$};
\draw[dashed] (2.5,1)--(2.5,0.8) node[midway,left=2pt] {$\epsilon_{f^\ast}$};
\draw[very thick] (2.7,0.4) arc(360:180:0.2);
\draw[->, very thick] (2.3,0.6)--(2.3,0.4) node[midway,left] {$f^\ast$};
\draw[very thick] (2.7,0.6)--(2.7,0.4);
\draw[very thick] (2.7,0.6) arc(0:180:0.2);
\fill (2.7,0.5) circle (0.6pt) node[right] {$\xi$};
\draw[help lines] (0')--(2') node[midway,left] {$ $};%
\draw[help lines] (3')--(1') node[midway,left] {$ $};%
\draw[help lines] (3')--(2') node[midway,left] {$ $};%
\end{tikzpicture}
\end{array},\\
&\mathrm{Tr_R}(\xi):=\epsilon_{f} \cdot (\xi \circ \id_{f^\ast} ) \cdot \eta_{f^\ast}=
\begin{array}{c}
\begin{tikzpicture}[scale=3]
\coordinate[label=above:$A$] (ct) at (2.5,0.4);
\coordinate[label=above:$B$] (cn) at (2.9,0);
\coordinate[label=right:$ $] (0') at (2,0) ;
\coordinate[label=above:$ $] (1') at (2,1) ;
\coordinate[label=left:$ $] (2')at (3,0) ;
\coordinate[label=below:$ $] (3') at (3,1) ;
\draw[help lines] (0')--(1');
\draw[dashed] (2.5,0)--(2.5,0.2) node[midway,left=2pt] {$\eta_{f^\ast}$};
\draw[dashed] (2.5,1)--(2.5,0.8) node[midway,left=2pt] {$\epsilon_f$};
\draw[very thick] (2.7,0.4) arc(360:180:0.2);
\draw[very thick] (2.3,0.6)--(2.3,0.4);
\draw[->, very thick] (2.7,0.6)--(2.7,0.4)  node[midway,right] {$f^\ast$};
\draw[very thick] (2.7,0.6) arc(0:180:0.2);
\fill (2.3,0.5) circle (0.6pt) node[left] {$\xi$};
\draw[help lines] (0')--(2') node[midway,left] {$ $};%
\draw[help lines] (3')--(1') node[midway,left] {$ $};%
\draw[help lines] (3')--(2') node[midway,left] {$ $};%
\end{tikzpicture}
\end{array}
\end{align}
One can further define the back 2-spherical trace $\mathrm{Tr_B}(\xi)$ and the front 2-spherical trace $\mathrm{Tr_F}(\xi)$ as
\begin{align}
\label{eq:back-trace}
\mathrm{Tr_B}(\xi):= \mathrm{Tr_R}(e_B \circ (\xi \ot B^\star))=\mathrm{Tr_L} ((\xi \ot A^\star) \circ i_{A^\star}).\\
\mathrm{Tr_F}(\xi):= \mathrm{Tr_R}(e_{B^\star} \circ (B^\star \ot \xi)=\mathrm{Tr_L} ((A^\star \ot \xi) \circ i_{A}).
\end{align}
Taking the right planar trace $\mathrm{Tr_R}(e_B \circ (\xi \ot B^\star))$ as example,  it can be graphically expressed as
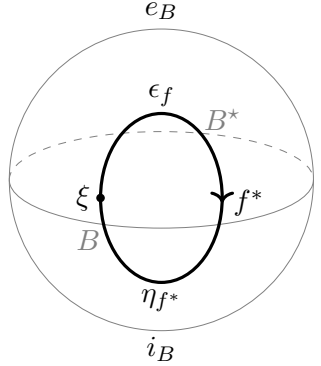
\begin{align*}
\begin{array}{c}
\begin{tikzpicture}[scale=0.8]
\draw[help lines] (5,1.5) circle (2.5) node[above=18pt] {$ $};
\draw[help lines] (7.5,1.5) arc (0:-180:2.5 and 0.8);
\node[help lines] at (3.8,0.5) {$B$};
\node[help lines] at (6,2.5) {$B^\star$};
\node at (5,-0.5) {$\eta_{f^\ast}$};
\node at (5,2.9) {$\epsilon_f$};
\node at (5,4.3) {$e_B$};
\node at (5,-1.3) {$i_B$};
\draw[help lines,dashed] (2.5,1.5) arc (180:0:2.5 and 0.8);
\draw[very thick] (5,1.2) ellipse (1 and 1.4);
\fill (4,1.2) circle (2pt) node[left] {$\xi$};
\draw[->, very thick] (6,1.201)--(6,1.119) node[right] {$f^\ast$};
\end{tikzpicture}
\end{array}
,
\end{align*}
i.e. evaluated by first stacking $B^\star$ from back on $\mathrm{Tr_R} (\xi): \id_B {\rightarrow} \id_B$ and then composing $i_B$ and $e_B$ from bottom and top.

In a spherical fusion 2-category, the back and front 2-spherical trace must agree, hence called 2-spherical trace, i.e. $\mathrm{Tr_F}(\xi)=\mathrm{Tr_B}(\xi) =: \mathrm{Tr}(\xi)$. The quantum dimension of a 1-morphism $f: A\to B$ is defined as $\dim(f) := \mathrm{Tr}(\id_f)$, and the quantum dimension of an object $A$ is defined as $\dim(A):= \dim(\id_A)$, where $\id_A$ is the identity 1-morphism.

\subsection{Normalized sections and retractions}
\label{sec:nor_sec_ret}

For 1-morphisms $f,g: A \rightarrow B$ in a spherical fusion 2-category, we can define the pairing $[ \cdot \ ,\ \cdot ]: \Hom(f,g) \times \Hom(g,f)\rightarrow k$ as
\begin{align}
    [ \rho, \xi ] :=\mathrm{Tr}(\rho \cdot \xi)=\mathrm{Tr}(\xi\cdot \rho),
\end{align}
where $\rho \in \Hom(f,g)$ and $\xi \in \Hom(g,f)$.

Let $(\rho, \xi)$ and $(\rho', \xi')$ be two pairs of section and retraction maps in the direct sum decomposition $A = \oplus X$.  The pairing $[ \rho', \xi ]=\mathrm{Tr}(\rho' \cdot \xi)=  \delta^{\xi'}_{ \xi} C \dim(X)$, where $C$ is a complex number. Then We introduce the normalized retraction and section maps in bra-ket notations as
\begin{align}
 \mid {\xi} \rangle & := \frac{\xi}{\sqrt{C \dim(X)}} &   \langle \xi \mid &:= \frac{\rho}{\sqrt{C \dim(X)}},
\end{align}
whose pairing is denoted as 
\begin{align*}
 \langle {\xi}' \mid {\xi} \rangle :=  [ \cll {\xi'}, \cl{\xi} ] =
\frac{1}{C \dim(X)} [ {\rho}', {\xi} ] = \delta^{\xi'}_\xi   
\end{align*}
Note that $\cl {\xi} $ can be considered as a basis of $\Hom(A, X)$, while $\cll {\xi} $ is a basis of $\Hom((X, A)$.

\subsection{The calculation of 10j-symbol}

As shown in eqn. \eqref{eq.sd}, eqn. \eqref{eq.ld}, and fig. \ref{eq.10j}, 
the calculation of 10j-symbol involves the normalized retractions $\mid {\zeta^1} \rangle, \mid {\zeta^2} \rangle, \mid {\zeta^3} \rangle, \mid {\zeta^4} \rangle$, and $\mid {\zeta^5} \rangle$, where ${\zeta^1}, {\zeta^2}, {\zeta^3}, {\zeta^4}$, and ${\zeta^5}$ are the corresponding retractions.  The bimodule maps $\mid {Z} \rangle$ and $\mid {YWXJ} \rangle$ are the compositions of the normalized retractions as shown in the figure and the equations.

With the spherical structure, we can do some surgery depicted as
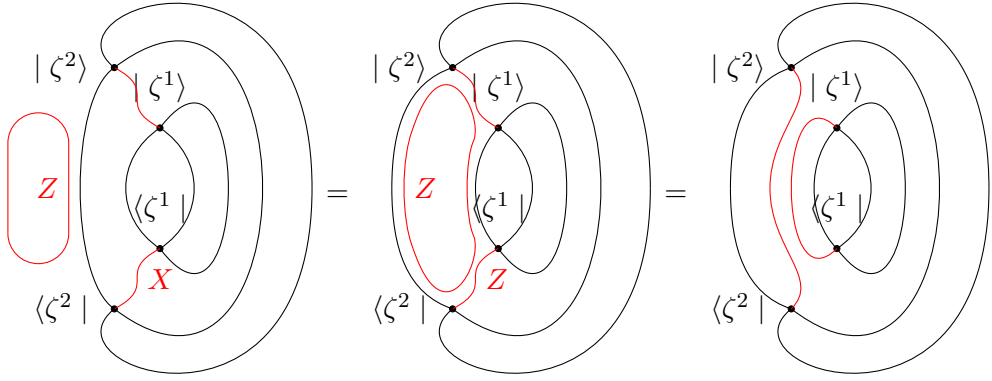
\begin{align*}
\begin{array}{c}
\begin{tikzpicture}[scale=0.4]
\draw[red] (-5,1.5) arc(180:0:1);
\draw[red] (-5,1.5)--(-5,-1.5);
\draw[red] (-3,1.5)--(-3,-1.5)node[midway,left]{$Z$};
\draw[red] (-5,-1.5) arc(-180:0:1);
\draw [](0,2).. controls (-1.5,1) and (-1.5,-1) ..(0,-2) node[midway,right]{$ $};
\draw [](0,2).. controls (1.5,1) and (1.5,-1) ..(0,-2) node[midway,right]{$ $};
\filldraw[] (0,2) circle (0.1) node[above=0.2] {$\mid {\zeta^1} \rangle$};
\filldraw[] (0,-2) circle (0.1) node[above=0.2] {$\langle {\zeta^1} \mid$};
\filldraw[] (-1.5,4) circle (0.1) node[left=0.2] {$\mid {\zeta^2} \rangle$};
\filldraw[] (-1.5,-4) circle (0.1) node[left=0.2] {$\langle {\zeta^2} \mid$};
\draw [](0,2).. controls (3,6)  and (3,-6) ..(0,-2);
\draw [red](0,2).. controls (-1.5,2.8)  and (0,3.2) ..(-1.5,4) node[midway,right]{$ $};
\draw [red](0,-2).. controls (-1.5,-2.8)  and (0,-3.2) ..(-1.5,-4) node[midway,right]{$X$};
\draw [](-1.5,4).. controls (5,9)  and (5,-9) ..(-1.5,-4);
\draw [](-1.5,4).. controls (-4,6)  and (5,9) ..(5,0);
\draw [](-1.5,-4).. controls (-4,-6)  and (5,-9) ..(5,0);
\draw [](-1.5,4).. controls (-3,3)  and (-3,-3) ..(-1.5,-4);
\end{tikzpicture}
\end{array}
=
\begin{array}{c}
\begin{tikzpicture}[scale=0.4]
\draw [](0,2).. controls (-1,1) and (-1,-1) ..(0,-2) node[midway,left]{$ $};
\draw [](0,2).. controls (1.5,1) and (1.5,-1) ..(0,-2) node[midway,left]{$ $};
\filldraw[] (0,2) circle (0.1) node[above=0.2] {$\mid {\zeta^1} \rangle$};
\filldraw[] (0,-2) circle (0.1) node[above=0.2] {$\langle {\zeta^1} \mid$};
\filldraw[] (-1.5,4) circle (0.1) node[left=0.2] {$\mid {\zeta^2} \rangle$};
\filldraw[] (-1.5,-4) circle (0.1) node[left=0.2] {$\langle {\zeta^2} \mid$};
\draw [](0,2).. controls (3,6)  and (3,-6) ..(0,-2);
\draw [red](0,2).. controls (-1.5,2.8)  and (0,3.2) ..(-1.5,4) node[midway,right]{$ $};
\draw [red](0,-2).. controls (-1.5,-2.8)  and (0,-3.2) ..(-1.5,-4) node[midway,right]{$Z$};
\draw [](-1.5,4).. controls (5,9)  and (5,-9) ..(-1.5,-4);
\draw [](-1.5,4).. controls (-4,6)  and (5,9) ..(5,0);
\draw [](-1.5,-4).. controls (-4,-6)  and (5,-9) ..(5,0);
\draw [](-1.5,4).. controls (-3,3.5)  and (-3.5,2) ..(-3.5,0);
\draw [](-1.5,-4).. controls (-3,-3.5)  and (-3.5,-2) ..(-3.5,0);
\draw [red](-0.8,1.6).. controls (-1.1,1) and (-1.1,-1) ..(-0.8,-1.6) node[midway,left]{$ $};
\draw [red](-0.8,1.6).. controls (-0.5,2.5) and (-1.8,5) ..(-2.8,2) node[midway,left]{$ $};
\draw [red](-0.8,-1.6).. controls (-0.5,-2.5) and (-1.8,-5) ..(-2.8,-2) node[midway,left]{$ $};
\draw [red](-2.8,2).. controls (-3.2,1) and (-3.2,-1) ..(-2.8,-2) node[midway,right]{$Z$};
\end{tikzpicture}
\end{array}
=
\begin{array}{c}
\begin{tikzpicture}[scale=0.4]
\draw [](0,2).. controls (-1,1) and (-1,-1) ..(0,-2) node[midway,left]{$ $};
\draw [](0,2).. controls (1.5,1) and (1.5,-1) ..(0,-2) node[midway,left]{$ $};
\filldraw[] (0,2) circle (0.1) node[above=0.2] {$\mid {\zeta^1} \rangle$};
\filldraw[] (0,-2) circle (0.1) node[above=0.2] {$\langle {\zeta^1} \mid$};
\filldraw[] (-1.5,4) circle (0.1) node[left=0.2] {$\mid {\zeta^2} \rangle$};
\filldraw[] (-1.5,-4) circle (0.1) node[left=0.2] {$\langle {\zeta^2} \mid$};
\draw [](0,2).. controls (3,6)  and (3,-6) ..(0,-2);
\draw [](-1.5,4).. controls (5,9)  and (5,-9) ..(-1.5,-4);
\draw [](-1.5,4).. controls (-4,6)  and (5,9) ..(5,0);
\draw [](-1.5,-4).. controls (-4,-6)  and (5,-9) ..(5,0);
\draw [](-1.5,4).. controls (-3,3.5)  and (-3.5,2) ..(-3.5,0);
\draw [](-1.5,-4).. controls (-3,-3.5)  and (-3.5,-2) ..(-3.5,0);
\draw [red](0,2).. controls (-2,4)  and (-2,-4) ..(0,-2);
\draw [red](-1.5,4).. controls (-0.5,3)  and (-2.2,2) ..(-2.2,0);
\draw [red](-1.5,-4).. controls (-0.5,-3)  and (-2.2,-2) ..(-2.2,0);
\end{tikzpicture}
\end{array}
\end{align*}
and get that $\dim(Z) \langle Z' \mid Z \rangle=\delta^Z_{Z'} \dim(Z) \langle {\zeta'^1} \ot {\zeta'^2} \mid {\zeta^1} \ot {\zeta^2} \rangle=\delta^Z_{Z'}\delta^{{\zeta'^1}}_{\zeta^1}\delta^{{\zeta'^2}}_{{\zeta^2}}$, where $ {\zeta^1} \ot {\zeta^2} $ are the juxtaposition of $ {\zeta^2} $ and $ {\zeta^1} $ as shown below
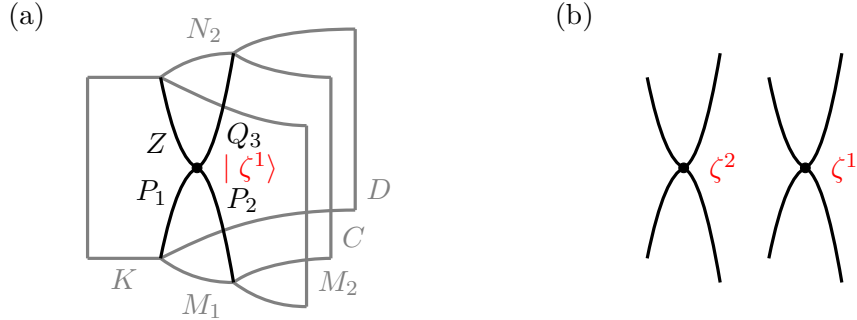
\begin{align*}
\centering
\begin{tikzpicture}[scale=1.6]
\node at (-0.5,2) {(a)};
\node at (4,2) {(b)};
\draw[help lines, very thick] (0,0)--(0.6,0) node[midway,below] {$K$};
\draw [help lines, very thick](0.6,0).. controls (1,0.2) and (1.4,0.4) ..(2.2,0.4) node[above right] {$D$};
\draw [help lines, very thick](0.6,0).. controls (0.8,-0.15) and (1.0,-0.2) ..(1.2,-0.2) node[below left] {$M_1$};
\draw [help lines, very thick](1.2,-0.2).. controls (1.4,-0.35) and (1.6,-0.4) ..(1.8,-0.4) node[above right] {$M_2$};
\draw [help lines, very thick](1.2,-0.2).. controls (1.4,-0.05) and (1.8,0) ..(2,0) node[above right] {$C$};
\draw[help lines, very thick] (0,0)--(0,1.5) node[midway,below] {$ $};
\draw[help lines, very thick] (0,1.5)--(0.6,1.5)
node[midway,below] {$ $};
\draw[help lines, very thick] (1.8,-0.4)--(1.8,1.1);
\draw[help lines, very thick] (2,0)--(2,1.5);
\draw[help lines, very thick] (2.2,0.4)--(2.2,1.9);
\draw [help lines, very thick](0.6,1.5).. controls (1,1.3) and (1.4,1.1) ..(1.8,1.1) node[above left] {$ $};
\draw [help lines, very thick](0.6,1.5).. controls (0.8,1.65) and (1.0,1.7) ..(1.2,1.7) node[above left] {$N_2$};
\draw [help lines, very thick](1.2,1.7).. controls (1.4,1.55) and (1.8,1.5) ..(2,1.5) node[above left] {$ $};
\draw [help lines, very thick](1.2,1.7).. controls (1.4,1.85) and (1.8,1.9) ..(2.2,1.9) node[above left] {$ $};
\draw [very thick](0.6,0).. controls (0.7,0.5) and (0.8,0.7) ..(0.9,0.75) node[midway,left] {$P_1$};
\draw [very thick](1.2,-0.2).. controls (1.1,0.4) and (1.0,0.7) ..(0.9,0.75) node[midway,right] {$P_2$};
\draw [very thick](0.6,1.5).. controls (0.7,1) and (0.8,0.8) ..(0.9,0.75) node[midway,left] {$Z$};
\draw [very thick](1.2,1.7).. controls (1.1,1.1) and (1.0,0.8) ..(0.9,0.75) node[midway,right] {$Q_3$};
\filldraw[] (0.9,0.75) circle (0.04) node[right=0.2,red] {$\mid {\zeta^1} \rangle$};
\draw [very thick](4.6,0).. controls (4.7,0.5) and (4.8,0.7) ..(4.9,0.75) node[midway,left] {$ $};
\draw [very thick](5.2,-0.2).. controls (5.1,0.4) and (5.0,0.7) ..(4.9,0.75) node[midway,right] {$ $};
\draw [very thick](4.6,1.5).. controls (4.7,1) and (4.8,0.8) ..(4.9,0.75) node[midway,left] {$ $};
\draw [very thick](5.2,1.7).. controls (5.1,1.1) and (5.0,0.8) ..(4.9,0.75) node[midway,right] {$ $};
\filldraw[] (4.9,0.75) circle (0.04) node[right=0.2,red] {$ {\zeta^2}$};
\draw [very thick](5.6,0).. controls (5.7,0.5) and (5.8,0.7) ..(5.9,0.75) node[midway,left] {$ $};
\draw [very thick](6.2,-0.2).. controls (6.1,0.4) and (6.0,0.7) ..(5.9,0.75) node[midway,right] {$ $};
\draw [very thick](5.6,1.5).. controls (5.7,1) and (5.8,0.8) ..(5.9,0.75) node[midway,left] {$ $};
\draw [very thick](6.2,1.7).. controls (6.1,1.1) and (6.0,0.8) ..(5.9,0.75) node[midway,right] {$ $};
\filldraw[] (5.9,0.75) circle (0.04) node[right=0.2,red] {$ {\zeta^1} $};
\end{tikzpicture}
\end{align*}
Consequently, we can define an invertible linear map
\begin{equation}
T=\sum_{Z} \mid {Z} \rangle \langle {\zeta^1}\ot {\zeta^2} \mid
\end{equation}
from $V_{{\zeta^1}\ot {\zeta^2}}$, the vector space spanned by ${\zeta^1}\ot {\zeta^2}$, to $V_{{Z}}$, the vector space spanned by $\mid Z \rangle$.  And its inverse reads
\begin{align}
\label{eq:T-1}
T^{-1}=&\sum_{Z} \dim(Z) \mid {\zeta^1}\ot {\zeta^2} \rangle \langle {Z} \mid,
\end{align}
which satisfies $T^{-1}T=\id_{V_{{\zeta^1} \ot {\zeta^2}}}$ and $TT^{-1}=\id_{V_{Z}}$.

Similarly, we have $\langle {\zeta'^3} \ot {\zeta'^4} \ot {\zeta'^5} \mid {\zeta^3} \ot {\zeta^4} \ot {\zeta^5} \rangle=\delta^{{\zeta'^3}}_{{\zeta^3}}\delta^{{\zeta'^4}}_{{\zeta^4}}\delta^{{\zeta'^5}}_{{\zeta^5}}$.
We can thus define another linear map
\begin{equation}
\label{eq:S}
S=\sum_{UYWXJ} \mid {YWXJ} \rangle \langle {\zeta^3} \ot {\zeta^4} \ot {\zeta^5} \mid,
\end{equation}
from $V_{{\zeta^3}\ot {\zeta^4} \ot {\zeta^5}}$ to $V_{{YWXJ}} = V_{{Z}}$, 
and its right inverse
\begin{equation}
S^{-1}=\sum_{UYWXJ} \dim(X)D^{WYU}_{J} \mid {\zeta^3} \ot {\zeta^4} \ot {\zeta^5} \rangle \langle {YWXJ} \mid
\end{equation}
satisfies $SS^{-1}=\id_{V_{{Z}}}$ and $\sum_{WYUJ} \langle {Y'W'X'J'} \mid {YWXJ} \rangle=\sum_{WYUJ} \delta^W_{W'}\delta^Y_{Y'}\delta^J_{J'} D^{WYU}_{J} \langle X' \mid X \rangle=\langle X' \mid X \rangle$ ($\mid X \rangle$ is defined in the same way as $\mid Z \rangle$), where
\begin{equation}
D^{WYU}_{J}=\frac{\dim(W) \dim(Y)}{ \dim(U) \dim(J) \dim(\mathrm{End}_{\Sigma \cB}(J)) n(J)},
\end{equation}
$n(J)$ is the number of equivalence classes of simple objects in the connected component of object $J$, $\dim(\mathrm{End}_{\Sigma \cB}(J))$ is the dimension of the fusion 1-category $\Hom(J,J)$, and $U$ is a simple 1-morphism in decomposition $\Hom(Y\circ W,Q_2 \circ Q_3)=\oplus_{U}  \Hom(Y\circ W,U) \circ \Hom(U,Q_2 \circ Q_3)$.

As a result, we can introduce a linear map $G = T^{-1} S$ from $V_{{\zeta^3}\ot {\zeta^4} \ot {\zeta^5}}$ to $V_{{\zeta^1} \ot {\zeta^2}}$.  According to eqn. \eqref{eq:T-1}, \eqref{eq:S}, and \eqref{eq:10j-G}, we have
\begin{align}
    G=T^{-1}S=&\sum_{UYWXZJ} \dim(Z) \mid {\zeta^1}\ot {\zeta^2} \rangle \langle {Z} \mid {YWXJ} \rangle \langle {\zeta^3} \ot {\zeta^4} \ot {\zeta^5} \mid\\ \nonumber
    =&\sum_{UYWXZJ} G_{Z}^{YWXJ} \mid {\zeta^1}\ot {\zeta^2} \rangle \langle {\zeta^3} \ot {\zeta^4} \ot {\zeta^5} \mid,
\end{align}
i.e. the 10j-symbol $G_{Z}^{YWXJ}$ is just the matrix element of the linear map $G$. Though the dimension of the vector spaces $V_{{\zeta^1} \ot {\zeta^2}}$ and $V_{{\zeta^3}\ot {\zeta^4} \ot {\zeta^5}}$ are in general different, the dimension of $V_{{\zeta^1} \ot {\zeta^2}}$, $V_{{Z}}$ and $V_{{YWXJ}}$ are same and called the dimension of the 10j-symbol.

Since $T^{-1}SS^{-1}T=\id_{V_{{\zeta^1} \ot {\zeta^2}}}$, it is clear that the right inverse of $G$ is given by
\begin{align}
G^{-1}=S^{-1}T=&\sum_{UYWXZ} \dim(X)D^{WYU}_{J} \mid {\zeta^3} \ot {\zeta^4} \ot {\zeta^5} \rangle \langle {YWXJ} \mid {Z} \rangle \langle {\zeta^1}\ot {\zeta^2} \mid \\ \nonumber
=&\sum_{UYWXZ} D^{WYU}_{J} (G^{-1})^{Z}_{YWXJ} \mid {\zeta^3} \ot {\zeta^4} \ot {\zeta^5} \rangle \langle {\zeta^1}\ot {\zeta^2} \mid.
\end{align}
Please note that $G_{Z}^{YWXJ}$ and $(G^{-1})^{Z}_{YWXJ}$ characterize the transformation between the two bases $\mid {YWXJ} \rangle$ and $\mid Z \rangle$.  And $G$ is only right invertible because the basis $\mid {YWXJ} \rangle$ is over-complete.
\section{Fusion 2-category $\Sigma \mathrm{sVec}$} \label{sVecbasic}
In this section, we will give a brief introduction of a spherical fusion 2-category $\Sigma \mathrm{sVec}$, the condensation completion of the braided fusion 1-category $\mathrm{sVec}$. $\mathrm{sVec}$ is the category of finite dimensional super vector spaces, which consists of the following data:
\begin{itemize}
    \item Two simple objects: $\one$, the one-dimensional vector space of grade 0, and $f$, the one-dimensional vector space of grade 1. They can be regarded as boson and fermion living on the surface of a 3+1D topological order, respectively.
    \item Tensor product: the tensor product of graded vector spaces.
    \item  Fusion rule:
        \begin{align}
            \one \ot \one = \one, \quad \one \ot f=f, \quad f\ot f= \one.
        \end{align}
    \item Trivial associator, i.e. $x\ot (y \ot z) = (x \ot y) \ot z$ for $x, y, z \in  \{\one, f\}$.
    \item The braiding: trivial except $c_{f, f}=-1$, which is consistent with fermionic statistics.
\end{itemize}
In the calculation, we will choose a basis for each objects in $\mathrm{sVec}$ and take the following nomenclatures
\begin{itemize}
    \item The basis of $\one$: $\{\cl 0\}$
    \item The basis of $f$: $\{\cl 1\}$
    \item The basis of $\A := \one \oplus f$: $\{\cl 0, \cl 1\}$
    \item The basis of $B \ot C$: $\{\cl b \cl c\}$ or simply $\{ \cl{bc} \}$, where $\{\cl b\}$ is the basis of $B$, and $\{\cl c\}$ is the basis of $C$.  For example, the basis of $\A \ot \A$ is denoted as $\{\cl 0 \cl 0, \cl 0 \cl 1, \cl 1 \cl 0, \cl 1 \cl 1\}$ or just $\{ \cl {00}, \cl {01}, \cl {10}, \cl {11} \}$.
\end{itemize}

$\Sigma \mathrm{sVec}$ is constructed following the definition in Sec. \ref{sec:sigmaB} and will be illustrated in detail below.

\subsection{Objects in $\Sigma \mathrm{sVec}$}
Objects in $\Sigma \mathrm{sVec}$ are separable algebras in $\mathrm{sVec}$.  There are two simple separable algebras in $\mathrm{sVec}$, the trivial algebra $\1:=(\one, m_\one: \one \ot \one \rightarrow \one)$ and a non-trivial algebra $(\A \equiv \one \oplus f, m_\A: \A \ot \A \rightarrow \A)$, where $m_\one$ is given by
\begin{align}
    \cl 0 \bullet \cl 0 = \cl 0,
\end{align}
and $m_\A$ is given by
\begin{align}
\cl a \bullet \cl b =\cl {a+b},
\end{align}
for $a, b \in \{0, 1\}$.  Please note that the addition within the brackets is always interpreted modulo 2.  In the following, we will denote the second algebra as $\A$ for simplicity.

All the other separable algebras in $\mathrm{sVec}$ are (Morita) equivalent to either $\1$ or $\A$~\cite{ostrik2003}.  For example, $\A\ot \A \cong \1$, which will be demonstrated in Sec. \ref{sec:mor_equi_1_AA}.  As a consequence, there are only two equivalence classes of the simple objects in $\Sigma \mathrm{sVec}$, and we take $\1$ and $\A$ as the representative objects. 

\subsection{1-morphisms in $\Sigma \mathrm{sVec}$}
1-morphisms in $\Sigma \mathrm{sVec}$ are bimodules in $\mathrm{sVec}$. For example, given two arbitrary objects $C$ and $B$ in $\Sigma \mathrm{sVec}$, i.e. two separable algebras in $\mathrm{sVec}$, a 1-morphism $M$ in $\Hom(B, C)$ is a $C$-$B$-bimodule $_C M_B$ in $\mathrm{sVec}$.

In general, a $C$-$B$-bimodule in a category $\cB$ can be regarded as a left $C \ot B^{rev}$-module (equivalent to a $C \ot B^{rev}$-$\1$-bimodule) , where $B^{rev} \equiv (B, m_B^{rev})$ with multiplication $m_B^{rev} = m_B \cdot c_{B, B}$ is also a separable algebra.  Therefore, the problem of finding simple bimodules reduces to the problem of finding simple left modules, which can be done by noting that all of the simple left modules of a separable algebra $D$ in $\cB$ can be realized by direct summands of free left-modules $D \ot x$ for all of the simple objects $x$ in $\cB$.

\begin{example}[Simple left modules of $\A \ot \A$]
\label{ex:left_mod_AA}
As mentioned above, the multiplication $m_A$ of $\A$ is given by
\begin{align}
\cl a \bullet \cl b =\cl {a+b}.
\end{align}
Then we can define an algebra $\A \ot \A$ with multiplication $(m_A \otimes m_A) \cdot (id_A \ot c_{A, A} \ot id_A)$, or simply
\begin{align}
\label{eq:mul_AA}
\cl a\cl b \bullet \cl c\cl d=(-)^{bc}\cl{a+c}\cl{b+d}.
\end{align}

$\A \ot \A$ is obvious a left $\A\ot \A$-modules with the action given by the algebra multiplication, and can be decomposed as ${\A \ot \A} = {V} \oplus {V'}$, where $V := (\A, l_V)$ and $V' := (\A, l_{V'})$ are two simple left $\A\ot \A$-modules with the action 
\begin{align*}
    \centering\begin{tabular}{c|cc}
        $l_V$ & $\cl 0$ & $\cl 1$\\
        \hline
        $\cl0\cl0 \act$ & $\cl0$ & $\cl1$\\
        $\cl0\cl1 \act$ & $\cl1$ & $\cl0$\\
        $\cl1\cl0 \act$ & $\ii\cl1$ & $-\ii\cl0$\\
        $\cl1\cl1 \act$ & $-\ii\cl0$ & $\ii\cl1$
    \end{tabular} & &
    \begin{tabular}{c|cc}
        $l_{V'} $ & $\cl 0$ & $\cl 1$\\
        \hline
        $\cl0\cl0 \act$ & $\cl0$ & $\cl1$\\
        $\cl0\cl1 \act$ & $\cl1$ & $\cl0$\\
        $\cl1\cl0 \act$ & $-\ii\cl1$ & $\ii\cl0$\\
        $\cl1\cl1 \act$ & $\ii\cl0$ & $-\ii\cl1$
    \end{tabular}
\end{align*}
and the section maps
\begin{align*}
    V & \to \A \ot \A & V' & \to \A \ot \A \\
    \cl0 &\mapsto \cl 0\cl 0+\ii\cl 1\cl1  &  \cl0 & \mapsto \cl 0\cl 0-\ii\cl 1\cl1 \\
    \cl1 &\mapsto \cl0 \cl 1-\ii\cl 1\cl0  &  \cl1 & \mapsto \cl0 \cl 1+\ii\cl 1\cl0 .
\end{align*}
Please check the appendix \ref{sec:decom_AA} for details.

$V$ and $V'$ are not isomorphic to each other since one can not find a module map, a map $u$ that preserves both the $Z_2$ grading and the algebra action, between them.  However, there is an invertible module map from $V\otimes f$ to $V'$:
\begin{align*}
 \cl 0_V \cl1 &\mapsto \cl 1_{V'},\\
 \cl 1_V \cl1 &\mapsto \cl 0_{V'},
\end{align*}
where the subscript shows explicitly which module the basis belongs to.  Therefore, $V'$ is isomorphic to $V\ot f \equiv Vf$, and we take $V$ and $Vf$ as the two representative simple $\A \ot \A$-$\1$-bimodules in the calculation.  Following our nomenclatures, the basis of $Vf$ is denoted as $\{\cl {01}, \cl {11}\}$.  Since $V \cong V \ot \one$, we sometimes denote the basis of $V$ as $\{\cl {00}, \cl {10}\}$ for a unified notation with the basis of $Vf$.
\end{example}

\begin{example}
[Simple $\A$-$\A$-bimodules]
$\A$-$\A$-bimodules can be regarded as left $\A\ot \A^{rev}$-modules, where the multiplication of $\A^{rev}$ is given by
\begin{align}
\cl a \bullet \cl b =(-)^{ab} \cl {a+b}.
\end{align}
Therefore, the multiplication in $\A \ot \A^{rev}$ is just
\begin{align}
\cl a \cl b \bullet \cl c \cl d:= (-)^{b(c+d)} \cl {a+c} \cl {b+d}.
\end{align}

Following the same method, we found two left $\A\ot \A^{rev}$-modules $W:= (\A, l_W)$ and $W' := (\A, l_{W'})$ with the section maps
\begin{align*}
W & \to \A \ot \A^{rev} & W' & \to \A \ot \A^{rev} \\
\cl 0 &\mapsto \cl0 \cl0+ \cl1 \cl1 & \cl 0 &\mapsto \cl0 \cl0- \cl1 \cl1\\    
\cl 1 &\mapsto \cl0 \cl1+ \cl1 \cl0 & \cl 1 &\mapsto \cl0 \cl1- \cl1 \cl0
\end{align*}

Then we can rewrite $W$ and $W'$ as $\A$-$\A$-bimodules with bimodule actions listed below
\begin{align*}
    \centering\begin{tabular}{c|cccc}
        $W$ & $\cl 0$ & $\cl 1$\\
        \hline
        $\cl0 \act$ & $\cl0$ & $\cl1$\\
        $\cl1 \act$ & $\cl1$ & $\cl0$\\
        \hline
        $\ract \cl0$ & $\cl0$ & $\cl1$\\
        $\ract \cl1$ & $\cl1$ & $\cl0$
    \end{tabular}    
&&
\begin{tabular}{c|cc}
        $W'$ & $\cl 0$ & $\cl 1$\\
        \hline
        $\cl0 \act$ & $\cl0$ & $\cl1$\\
        $\cl1 \act$ & $-\cl1$ & $-\cl0$\\
        \hline
        $\ract \cl0$ & $\cl0$ & $\cl1$\\
        $\ract \cl1$ & $\cl1$ & $\cl0$
    \end{tabular}.
\end{align*}

Alternatively, since $\A$ is a separable algebra, $\A$ itself can be regarded as a simple $\A$-$\A$-bimodule with the action given by the multiplication of the algebra.  We can construct another bimodule $f\A := f \ot \A$, where the left action is given by $\A\act (f\A) \xrightarrow{c_{\A, f}} f \ot (\A \act \A) = f \ot (\A \bullet \A) \rightarrow f\A $.  It is clear that $W \cong \A$ and $W' \cong f\A$, hence we will choose $\A$ and $f\A$ as the two representative simple $\A$-$\A$-bimodules.  Following our nomenclatures, the basis of $f\A$ are denoted as $\{\cl {10}$, $\cl {11}\}$. For a unified notation, we sometimes denote the basis of $\A$ as $\{\cl {00}$, $\cl {01}\}$. 
\end{example}

With the above methods, we can fix the choice of representative simple 1-morphisms.  In the following calculation, we use only the 1-morphisms in $\Hom(B,C\ot D)$, for $B, C, D \in \{\1, \A\}$, and the corresponding representative bimodules are 
$\III$, $\IfI$, $\IAA$, $\AAI$, $\AAA$, $\AfAA$, $\AAV$, $\AAVf$ and $\AAAAA$, where $\AAAAA = \AAI  \ot  \AAA$.  Note that we have used the relation $\one \ot X = X \ot \one = X$.

\subsection{Composition of 1-morphisms}
The composition of 1-morphisms (bimodules) is given by the relative tensor product of modules
\begin{align}
    \circ: \Hom(B,C) \xt \Hom(C,D) &\to \Hom(B,D),\\
    (_C N_B, {}_D M_C)&\mapsto M\circ N:={}_D M_C \ot[C]{}_C N_B, \nonumber
\end{align}
where the relative tensor product $M\ot[C]N$ is given by a quotient map $M \ot N \to M\ot[C]N$ satisfying
\begin{align}
    (\cl m\ract \cl c ) \cl n = \cl m (\cl c  \act \cl n) \rightarrow \cl m \ot[C] \cl n,
\end{align}
$\forall m \in M$, $c \in C$, $n \in N$.  In case of no confusion, we will simplify $_D M_C \ot[C]{}_C N_B$ to $_D M \ot[C] N_B$.  Below we will give some examples of the composition of 1-morphisms, which are going to be used in the following calculations.

\begin{example}
[Composition of $_\1 V^{rev} _{\A\ot \A}$ and $\AAV$] $V^{rev}$ is a bimodule induced from $V$. It is a same vector space as $V$, and the right action on $V^{rev}$ is induced from the left action on $V$ through
\begin{align}
\cl d\ract\cl a\cl b:=\cl b\cl a \act \cl d = (-)^{(a+d)b} (\ii)^b\cl{a+b+d}.
\end{align}
Then the composition of $V^{rev}$ and $V$ reads
\begin{align}
    _\1 V^{rev} _{\A\ot \A} \circ \AAV = _\1 V^{rev}\ot[\A \ot \A] V_\1 = \III,
\end{align}
with the quotient map
\begin{align*}
    V^{rev}\ot V &\to \one\\
    \cl 0_{V^{rev}}\cl 0_V+i \cl 1_{V^{rev}}\cl 1_V &\mapsto \cl 0\\
    \cl 0_{V^{rev}}\cl 1_V+i \cl 1_{V^{rev}}\cl 0_V &\mapsto 0.
\end{align*}
The detailed calculation can be found in Appendix.~\ref{sec:VVTI}
\label{ex:V_V}
\end{example}

\begin{example}
[$\AAA \circ \AAA$]
$\AAA \ot[\A] \AAA = \AAA$ via the quotient map 
\begin{align*}
    \A \ot \A &\to \A \\
    \cl 0  \cl 0 + \cl 1 \cl 1 & \mapsto \cl 0 
    \\
    \cl 0  \cl 1 + \cl 1 \cl 0 & \mapsto \cl 1
    \\
    \cl 0  \cl 0 - \cl 1 \cl 1 & \mapsto 0 
    \\
    \cl 0  \cl 1 - \cl 1 \cl 0 & \mapsto 0
\end{align*}
In the subsequent discussion, we will implicitly omit basis vectors that map to 0 (for example the last two lines of the preceding equations) for the sake of brevity.
\end{example}

\begin{example}[$(\A \ot \A) \circ Vf$]
\label{ex:AA-comp-Vf}
    $(\A \ot \A) \circ Vf = Vf$ with quotient map given by the left action on $Vf$
    \begin{align*}
     \A \ot \A \ot Vf &\to Vf \\
     \cl0 \cl0 \cl{01}+\cl0 \cl1 \cl{11}+i\cl1 \cl0 \cl{11}+i\cl1 \cl1 \cl{01} &\mapsto \cl {01}\\
     \cl0 \cl0 \cl{11}+\cl0 \cl1 \cl{01}-i\cl1 \cl0 \cl{01}-i\cl1 \cl1 \cl{11} &\mapsto \cl {11}.
    \end{align*}
\end{example}

\begin{example}
\label{ex:AAVf-comp-Vf}
    $(\A \ot \A \ot Vf) \circ Vf = Vf \ot Vf$ with quotient map
\begin{align*}
 \A \ot \A \ot Vf \ot Vf &\to Vf \ot Vf\\
\cl0 \cl0 \cl{01} \cl{01}-\cl0 \cl1 \cl{01} \cl{11}-i\cl1 \cl0 \cl{01} \cl{11}+i\cl1 \cl1 \cl{01} \cl{01} &\mapsto \cl{01} \cl {01}\\
\cl0 \cl0 \cl{11} \cl{01}+\cl0 \cl1 \cl{11} \cl{11}+i\cl1 \cl0 \cl{11} \cl{11}+i\cl1 \cl1 \cl{11} \cl{01} &\mapsto \cl{11} \cl {01}\\
\cl0 \cl0 \cl{01} \cl{11}-\cl0 \cl1 \cl{01} \cl{01}+i\cl1 \cl0 \cl{01} \cl{01}-i\cl1 \cl1 \cl{01} \cl{11} &\mapsto \cl{01} \cl {11}\\
\cl0 \cl0 \cl {11} \cl{11}+\cl0 \cl1 \cl {11} \cl{01}-i\cl1 \cl0 \cl {11} \cl{01}-i\cl1 \cl1 \cl {11} \cl{11} &\mapsto \cl {11} \cl {11}.
\end{align*}
\end{example}

\begin{example} $\AAAAA \circ \AAI = \A \ot \A$ with quotient map
    \begin{align*}
    \A \ot \A \ot \A &\to \A \ot \A\\
    \cl 0 \cl 0 \cl 0 + \cl 0 \cl 1 \cl 1 &\mapsto \cl 0 \cl 0 \\
    \cl 1 \cl 0 \cl 0 + \cl 1 \cl 1 \cl 1 &\mapsto \cl 1 \cl 0 \\
    \cl 0 \cl 0 \cl 1 + \cl 0 \cl 1 \cl 0 &\mapsto \cl 0 \cl 1 \\
    \cl 1 \cl 0 \cl 1 + \cl 1 \cl 1 \cl 0 &\mapsto \cl 1 \cl 1.
    \end{align*}
\end{example}

\begin{example} $(\A \ot \AAI) \circ \AAI = \A \ot \A$ with quotient map
\begin{align*}
    \A \ot \A \ot \A &\to \A \ot \A\\
    \cl 0 \cl 0 \cl 0 + \cl 1 \cl 0 \cl 1 &\mapsto \cl 0 \cl 0 \\
    \cl 0 \cl 1 \cl 0 - \cl 1 \cl 1 \cl 1 &\mapsto \cl 1 \cl 0 \\
    \cl 0 \cl 0 \cl 1 + \cl 1 \cl 0 \cl 0 &\mapsto \cl 0 \cl 1 \\
    \cl 0 \cl 1 \cl 1 - \cl 1 \cl 1 \cl 0 &\mapsto \cl 1 \cl 1.
\end{align*}
    
\end{example}

\subsection{2-morphisms in $\Sigma \mathrm{sVec}$}

2-morphisms in $\Sigma \mathrm{sVec}$ are bimodule maps. For two arbitrary $C$-$B$-bimodules $_C M_B$ and $_C N_B$, a bimodule map is a linear map $u$ between the two vector spaces $M$ and $N$ satisfying
\begin{align}
\label{bmm}
&u(c\act m)=c \act u(m), \quad u(m \ract b)=u(m) \ract b,
\end{align}
$\forall c\in C,\quad b\in B,\quad m\in M \nonumber$.  For given bases of $M$ and $N$, the bimodule map can be expressed as a matrix, while the composition of bimodule maps is just matrix multiplication.  And it is obvious that the product of $u$ and any nonzero complex number $z$ is also a bimodule map.
\subsection{Morita equivalence of objects in $\Sigma \mathrm{sVec}$}
\label{sec:mor_equi_1_AA}

Two algebras $B$ and $C$ are Morita equivalent if and only if there exists an invertible bimodule $_B M_C$, or in other words, there is an invertible 1-morphism between $B$ and $C$.  Morita equivalence is of particular importance as it allows us to concentrate on a finite number of equivalent classes of objects, rather than an infinite number of objects in  $\Sigma \mathrm{sVec}$.

\begin{example}
    [$\A\ot \A$ is Morita equivalent to $\1$] We will show that $_{\A\ot \A} V _\one$ is invertible. We have shown that $_\1 V^{rev}\ot[\A\ot \A] V_\1 = \III$ in example \ref{ex:V_V}.  For $_{\A \ot \A} V\ot V^{rev} _{\A \ot \A}$, there is an invertible $\A \ot \A$-$\A \ot \A$-bimodule map $ V\ot V^{rev} \rightarrow \A\ot \A$ given by
\begin{align*}
    \cl0_V\cl0_{V^{rev}}& \mapsto \cl0\cl0 +\ii \cl1\cl1,\\
    \cl1_V\cl1_{V^{rev}}& \mapsto -\ii\cl0\cl0 -\cl1\cl1,\\
    \cl1_V\cl0_{V^{rev}} & \mapsto \cl0\cl1 -\ii \cl1\cl0,\\
    \cl0_V\cl1_{V^{rev}} & \mapsto -\ii\cl0\cl1 +\cl1\cl0.
\end{align*}
Therefore, we have $_{\A \ot \A} V\ot V^{rev} _{\A \ot \A} \cong _{\A \ot \A} \A\ot \A _{\A \ot \A}$, hence $_{\A\ot \A} V _\one$ is invertible, and $\A\ot \A$ is Morita equivalent to $\1$.
\end{example}

With the same approach, we can find that there are just two Morita equivalent classes of simple objects in $\mathrm{sVec}$, one is with $\1$, the other is with $\A$.  In the calculation of 10j-symbol, we only need consider the representative objects of these two classes, which are chosen as $\1$ and $\A$ respectively.

\subsection{Tensor product of bimodules}
Recall that for two arbitrary bimodules $_C N_B $ and $ _Z P_Y$, we can define their tensor product $N\ot P$, which has a natural structure of $C\ot Z$-$B\ot Y$-bimodule (see Sec. \ref{sec:ten_pro_bim}). In $\mathrm{sVec}$ case, the bimodule structure is given by
\begin{align}
    \cl c \cl z \act \cl n \cl p &:=(-)^{zn} \cl {c\act n} \cl {z \act p} \nonumber\\
    \cl n \cl p \ract \cl b \cl y &:=(-)^{bp} \cl {n \ract b} \cl {p \ract y}.
\end{align}
Since an object $B$ can be regarded as the trivial 1-morphism $_B B_B$ in $\Hom(B,B)$, the tensor product $ _D M_C \ot B$ can be defined as $_{D\ot B} M\ot B _{C\ot B}$.

\begin{example}
[Tensor product of $\AAV$ and $\AAA$]
As discussed above, $\AAV \ot \AAA = \tensor*[_{\A\ot \A \ot \A}]{(}{}V \ot \A\tensor*{)}{_\A} =: V\A$, where the left action is twisted by $c_{V,\A}$, while the right $\A$-action is untwisted and acted on $\A$ in $V \ot \A$. The action is expressed below
\begin{align}
    \label{tab.VA}
        \centering\begin{tabular}{c|cccc}
        $V\A$ & $\cl{00}$ & $\cl {01}$ & $\cl {10}$ & $\cl {11}$\\
        \hline
        $\cl{000} \act $& $\cl{00}$ & $\cl {01}$ & $\cl {10}$ & $\cl {11}$ \\
        $\cl{010} \act $& $\cl{10}$ & $\cl {11}$ & $\cl {00}$ & $\cl {01}$ \\
        $\cl{100} \act $& $i\cl{10}$ & $i\cl {11}$ & $-i\cl {00}$ & $-i\cl {01}$ \\
        $\cl{110} \act $& $-i\cl{00}$ & $-i\cl {01}$ & $i\cl {10}$ & $i\cl {11}$ \\
        $\cl{001} \act $& $\cl{01}$ & $\cl {00}$ & $-\cl {11}$ & $-\cl {10}$ \\
        $\cl{011} \act $& $\cl{11}$ & $\cl {10}$ & $-\cl {01}$ & $-\cl {00}$ \\
        $\cl{101} \act $& $i\cl{11}$ & $i\cl {10}$ & $i\cl {01}$ & $i\cl {00}$ \\
        $\cl{111} \act $& $-i\cl{01}$ & $-i\cl {00}$ & $-i\cl {11}$ & $-i\cl {10}$ \\
        \hline
        $\ract \cl{0} $& $\cl{00}$ & $\cl {01}$ & $\cl {10}$ & $\cl {11}$ \\
        $\ract \cl{1} $& $\cl{01}$ & $\cl {00}$ & $\cl {11}$ & $\cl {10}$
    \end{tabular}
\end{align}
\end{example}

\subsection{The retraction bimodule maps}

Recall that the retraction bimodule maps in the 
direct sum decomposition of $A\ot (B \ot C)$-$K$-bimodule $\Lambda_{A, B, C} \circ (Q\ot C)\circ P = \bigoplus (A\ot Y)\circ X$  plays crucial roles in the calculation of 10j-symbol, where $K, A, B, C, M, N \in\Sigma\cB_0$, $P\in \ho(K,M\ot C)$, $Q\in \ho(M, A\ot B)$, $X\in \ho(K,A\ot N)$, $Y\in \ho(N,B\ot C)$.

In the $\Sigma \mathrm{sVec}$ case, the representative objects are $\Sigma\mathrm{sVec}_0 = \{\1, \A\}$.  The representative 1-morphisms are chose as $\ho(\1, \1 \ot \1) = \{\III, \IfI\}$, $\ho(\1, \A \ot \1) = \ho(\1, \1 \ot \A) = \{\AAI\}$, $\ho(\A, \1 \ot \1) = \{\IAA\}$, $\ho(\A, \A \ot \1) = \ho(\A, \1 \ot \A) = \{\AAA, \AfAA\}$, $\ho(\1, \A \ot \A) = \{\AAV, \AAVf\}$ and $\ho(\A, \A \ot \A) = \{\AAAAA\}$. Note that the data of $\Sigma \mathrm{sVec}$ can be used to describe a $2+1D$ boundary of a $3+1D$ topological order.  The object $\A$ represents a Majorana chain, while the object $\1$ represents the trivial chain (or just nothing) on the 2+1D boundary.  The 1-morphisms are domain walls.  For example, $\III$ and $\IfI$ are domain walls between trivial chains (or just nothing), hence are just boson and fermion particle respectively. $\IAA$ and $\AAI$ are the domain wall between the Majorana chain and the trivial chain, i.e. the Majorana zero modes.  $\AAA$ and $\AfAA$ are the particles lived on the Majorana chain, where $\AfAA$ is $\AAA$ decorated by a fermion. Please note that $\AAA$ is NOT a Majorana zero mode.  We will show later that both $\AAA$ and $\AfAA$ have quantum dimension 1 instead of $\sqrt{2}$. Similarly, $\AAVf$ can be considered as $\AAV$ with a decorated fermion, and both of them are domain walls between vacuum and a double-Majorana-chain.  $\AAAAA$ is domain wall between a Majorana chain and a double-Majorana chain, hence a Majorana zero mode.

Since all the associators in $\mathrm{sVec}$ are trivial, the associator bimodule $\Lambda_{A,B,C}$ is just an identity and will be dropped in the following. Then the direct sum decomposition reduces to  $(Q\ot C)\circ P = \bigoplus (A\ot Y)\circ X$. Below, we will give an example on how to calculate the retraction maps.

\begin{example}
    [Retraction map in the decomposition of $(V\ot \A)\circ \A$]
In this example, we consider the retraction map in the direct sum decomposition of $(\AAV \ot \A)\circ \AAA = \bigoplus (\A \ot Y) \circ X$ for $Y \in \ho(\1, \A \ot \A) = \{\AAV, \AAVf\}$ and $X \in \ho(\A, \A \ot \1) = \{\AAA, \AfAA\}$. 
With the standard procedure, we have $(\AAV \ot A)\circ \AAA = V \otimes \A = V\A$ with the quotient map 
\begin{align*}
V \ot \A \ot \A &\to V\A \\
\cl0_V \cl0_\A \cl0_\A+\cl0_V \cl1_\A \cl1_\A &\mapsto \cl {00}\\
\cl0_V \cl0_\A \cl1_\A+\cl0_V \cl1_\A \cl0_\A &\mapsto \cl {01}\\
\cl1_V \cl0_\A \cl0_\A+\cl1_V \cl1_\A \cl1_\A &\mapsto \cl {10}\\
\cl1_V \cl0_\A \cl1_\A+\cl1_V \cl1_\A \cl0_\A &\mapsto \cl {11}.
\end{align*}
Then we consider the bimodule $(\A\ot \AAV)\circ \AAA =: \widetilde{V\A}$, which corresponds to $Y=\AAV$ and $X=\AAA$.  Similarly, 
we have $\widetilde{V\A}$ is the same vector space as $V\A$, but with different action, which is presented below
\begin{center}
    \begin{tabular}{c|cccc}
        ${\widetilde{V\A}}$ & $\cl{00}$ & $\cl {01}$ & $\cl {10}$ & $\cl {11}$\\
        \hline
        $\cl{000} \act$ & $\cl{00}$ & $\cl {01}$ & $\cl {10}$ & $\cl {11}$ \\
        $\cl{010} \act$ & $i\cl{10}$ & $i\cl {11}$ & $-i\cl {00}$ & $-i\cl {01}$ \\
        $\cl{100} \act$ & $\cl{01}$ & $\cl {00}$ & $-\cl {11}$ & $-\cl {10}$ \\
        $\cl{110} \act$ & $-i\cl{11}$ & $-i\cl {10}$ & $-i\cl {01}$ & $-i\cl {00}$ \\

        $\cl{001} \act$ & $\cl{10}$ & $\cl {11}$ & $\cl {00}$ & $\cl {01}$ \\
        $\cl{011} \act$ & $-i\cl{00}$ & $-i\cl {01}$ & $i\cl {10}$ & $i\cl {11}$ \\
        $\cl{101} \act$ & $-\cl{11}$ & $-\cl {10}$ & $\cl {01}$ & $\cl {00}$ \\
        $\cl{111} \act$ & $-i\cl{01}$ & $-i\cl {00}$ & $-i\cl {11}$ & $-i\cl {10}$ \\
        \hline
        $\ract \cl{0} $& $\cl{00}$ & $\cl {01}$ & $\cl {10}$ & $\cl {11}$ \\
        $\ract \cl{1} $& $\cl{01}$ & $\cl {00}$ & $\cl {11}$ & $\cl {10}$
    \end{tabular}
\end{center}
The quotient map is given by
\begin{align*}
 \A \ot V \ot \A &\to \widetilde{V\A} \\
\cl0_\A \cl0_V \cl0_\A+\cl1_\A \cl0_V \cl1_\A &\mapsto \cl {00}\\
\cl0_\A \cl0_V \cl1_\A+\cl1_\A \cl0_V \cl0_\A &\mapsto \cl {01}\\
\cl0_\A \cl1_V \cl0_\A-\cl1_\A \cl1_V \cl1_\A &\mapsto \cl {10}\\
\cl0_\A \cl1_V \cl1_\A-\cl1_\A \cl1_V \cl0_\A &\mapsto \cl {11}
\end{align*}
Since, there is an invertible bimodule map $\zeta$ from $V\A$ to $\widetilde{V\A}$ defined as
\begin{align*}
    \cl {00} \mapsto \frac{1}{\sqrt{2}}(\cl {00} +\cl {11}),\\
    \cl {01} \mapsto \frac{1}{\sqrt{2}}(\cl {01} +\cl {10}),\\
    \cl {10} \mapsto \frac{1}{\sqrt{2}}(-i\cl {01} +i\cl {10}),\\
    \cl {11} \mapsto \frac{1}{\sqrt{2}}(-i\cl {00} +i\cl {11}).\\ 
\end{align*}
We have the direct sum decomposition $(V\ot \A)\circ \A = (\A\ot V)\circ \A$ with $\zeta$ as the retraction map and its reverse $\zeta^{-1}$ as the section map.  Graphically, the retraction can be expressed as
\begin{equation}
\begin{tikzcd}
&
    \itk{
    \coordinate (a)  at (-3,4);
    \coordinate (b)  at (-1,4);
    \coordinate (c)  at (1,4);
    \coordinate (p) at (-2,3);
    \coordinate (x) at (-1,2);
    \coordinate (y)  at (1,2);
    \coordinate (q)  at (2,3);
    \coordinate (w)  at (0,3);
    \coordinate (k)  at (-1,0.5);
    \node[above] at (a) {$ $};
    \node[above] at (b) {$ $};
    \node[above] at (c) {$ $};
    \node[above] at (p) {$V$};
    \node[above] at (x) {$\A$};
    \node[below] at (k) {$ $};
    \draw (a) -- (p);
    \draw [dashed] (p) -- (x);
    \draw (b) -- (p);
    \draw (c) -- (x);
    \draw (x) -- (k);
    }
\arrow[r,rightarrow,"\text{retraction}"]
&
\itk{
    \coordinate (a1)  at (5,4);  
    \coordinate (b1)  at (7,4);
    \coordinate (c1)  at (9,4);
    \coordinate (p1) at (8,3);
    \coordinate (x1) at (7,2);
    \coordinate (y1)  at (9,2);
    \coordinate (q1)  at (10,3);
    \coordinate (w1)  at (8,3);
    \coordinate (k1)  at (7,0.5);
    \node[above] at (a1) {$ $};
    \node[above] at (b1) {$ $};
    \node[above] at (c1) {$ $};
    \node[above] at (p1) {$V$};
    \node[above] at (x1) {$\A$} ;
    \node[below] at (k1) {$ $};
    \draw (a1)--(x1);
    \draw (b1)--(p1);
    \draw (c1)--(p1);
    \draw [dashed] (p1)--(x1);
    \draw (x1)--(k1);
    }
\end{tikzcd}
\end{equation}

\end{example}

\subsection{The interchangers}

Another important bimodule map in our calculation is the interchanger $\phi_{\tensor[_C]{N}{_B}, \tensor[_Z]{P}{_Y}}$, which is given by
\begin{align*}
(N \ot Z)\circ(B\ot P) &\stackrel{\tl{c}_{B,Z;N, P}}{\longrightarrow} (N\circ B)\ot (Z\circ P) \cong N\ot P \nonumber\\
\cong (C\circ N)\ot(P\circ Y) &\stackrel{\tl{c}_{P,N;C, Y}^{-1}}{\longrightarrow} (C\ot P)\circ (N\ot Y),
\end{align*}
where the 2-morphism $\tilde{c}_{B, Z;N, P}$ is induced from the braiding $c_{B, Z}$ in $\mathrm{sVec}$ as shown in eqn. \eqref{eq:ctilde}.

As an example, we consider the interchanger $\phi_{\AAVf, \AAVf}$ given by
\begin{align*}
    \phi_{Vf, Vf} = \tl{c}_{Vf,Vf; \A\ot \A,\1}^{-1} \circ \theta \circ \tl{c}_{\1,\A \ot \A; Vf, Vf},
\end{align*}
where $\theta$ is the 2-isomorphism $Vf \ot ((\A\ot \A)\circ Vf) \cong ((\A\ot \A)\circ Vf)\ot Vf$.  The interchanger can be depicted as
\begin{equation}
\begin{tikzcd}
&
    \itk{
    \coordinate (a)  at (-3,4);
    \coordinate (b)  at (-1,4);
    \coordinate (c)  at (1,4);
    \coordinate (d)  at (3,4);
    \coordinate (p) at (-2,3);
    \coordinate (x) at (-1,2);
    \coordinate (z)  at (0,1);
    \coordinate (y)  at (1,2);
    \coordinate (q)  at (2,3);
    \coordinate (w)  at (0,3);
    \coordinate (k)  at (0,-0.5);
    \node[above] at (a) {$ $};
    \node[above] at (b) {$ $};
    \node[above] at (c) {$ $};
    \node[above] at (d) {$ $};
    \node[above] at (p) {$Vf$};
    \node[above left] at (y) {$Vf$};
    \node[above] at (z) {$ $};
    \node[below] at (k) {$ $};
    \draw (a) -- (p);
    \draw [dashed] (p) --(z) -- (k);
    \draw (b) -- (p);
    \draw (c) -- (y);
    \draw (d) -- (y);
    \draw[dashed] (y) -- (z);
} 
\arrow[r,rightarrow,"\text{Ic}"]
&
\itk{
    \coordinate (a)  at (-3,4);
    \coordinate (b)  at (-1,4);
    \coordinate (c)  at (1,4);
    \coordinate (d)  at (3,4);
    \coordinate (p) at (-2,3);
    \coordinate (x) at (-1,2);
    \coordinate (z)  at (0,1);
    \coordinate (y)  at (1,2);
    \coordinate (q)  at (2,3);
    \coordinate (w)  at (0,3);
    \coordinate (k)  at (0,-0.5);
    \node[above] at (a) {$ $};
    \node[above] at (b) {$ $};
    \node[above] at (c) {$ $};
    \node[above] at (d) {$ $};
    \node[above right] at (x) {$Vf$};
    \node[above] at (q) {$Vf$};
    \node[above] at (z) {$ $};
    \node[below] at (k) {$ $};
    \draw (a) -- (x);
    \draw [dashed] (x) --(z) -- (k);
    \draw (b) -- (x);
    \draw (c) -- (q);
    \draw (d) -- (q);
    \draw [dashed] (q) -- (z);
}
\label{itcexample}
\end{tikzcd}
\end{equation}
We start from $\tl{c}_{Vf, Vf; \A\ot \A,\1}$ and $\tl{c}_{\1, \A \ot \A; Vf, Vf}$, which are computed in the following examples.

\begin{example}
$\tl{c}_{Vf, Vf; \A\ot \A,\1}: (\A \ot \A \ot Vf) \circ Vf \rightarrow  ((\A \ot \A) \circ Vf) \ot Vf$ is induced from $c_{Vf, Vf}$ in $\mathrm{sVec}$, which is given by
\begin{align*}
    \cl {01}_{Vf} \cl{01}_{Vf} &\mapsto -\cl {01}_{Vf} \cl{01}_{Vf},\\
    \cl {01}_{Vf} \cl{11}_{Vf} &\mapsto \cl {11}_{Vf} \cl{01}_{Vf},\\
    \cl {11}_{Vf} \cl{01}_{Vf} &\mapsto \cl {01}_{Vf} \cl{11}_{Vf},\\
    \cl {11}_{Vf} \cl{11}_{Vf} &\mapsto \cl {11}_{Vf} \cl{11}_{Vf}.\\ 
\end{align*}

According to example \ref{ex:AA-comp-Vf} and \ref{ex:AAVf-comp-Vf}, we have $(\A \ot \A) \circ Vf = Vf$ and $(\A \ot \A \ot Vf) \circ Vf = Vf \ot Vf$.  Therefore, we have
\begin{align*}
\tl{c}_{Vf, Vf; \A\ot \A,\1}: Vf \ot Vf &\rightarrow Vf \ot Vf, \\
    \cl {01} \cl{01} &\mapsto -\cl {01} \cl{01},\\
    \cl {01} \cl{11} &\mapsto \cl {11} \cl{01},\\
    \cl {11} \cl{01} &\mapsto \cl {01} \cl{11},\\
    \cl {11} \cl{11} &\mapsto \cl {11} \cl{11}.\\ 
\end{align*}
\end{example}

\begin{example}[$\tl{c}_{\1, \A \ot \A; Vf, Vf}$]
It is clear that ${c}_{\1, \A \ot \A}=\id_{ \A \ot \A }$. According to example \ref{ex:AA-comp-Vf}, we have $\tl{c}_{\1, \A \ot \A; Vf, Vf}=\id_{Vf \ot Vf}$.
\end{example}

With these $\tilde{c}$, we can calculate the interchange bimodule maps $\phi_{Vf, Vf}$. Since $(\A \ot \A) \circ Vf = Vf$ (example \ref{ex:AA-comp-Vf}), we have $\theta=\id_{Vf \ot Vf}$.  With $\tl{c}_{\1, \A \ot \A; Vf, Vf}=\id_{Vf \ot Vf}$, we have $\phi_{Vf, Vf}=\tl{c}_{Vf, Vf; \A\ot \A,\1}^{-1}$, hence
\begin{align*}
\phi_{Vf, Vf}: Vf \ot Vf &\rightarrow Vf \ot Vf, \\
    \cl {01} \cl{01} &\mapsto -\cl {01} \cl{01},\\
    \cl {01} \cl{11} &\mapsto \cl {11} \cl{01},\\
    \cl {11} \cl{01} &\mapsto \cl {01} \cl{11},\\
    \cl {11} \cl{11} &\mapsto \cl {11} \cl{11}.\\ 
\end{align*}

\subsection{Quantum dimension}

The quantum dimension of a 1-morphism $f$ of a spherical fusion 2-category is defined as $\dim(f) := \mathrm{Tr}(\id_f)$, where the 2-spherical trace $\mathrm{Tr}(\xi)$ of a 2-morphism $\xi$ is defined in eqn. \eqref{eq:back-trace}. Below is an example on the quantum dimension of the 1-morphism $\AAI$ in $\Sigma \mathrm{sVec}$.

\begin{example}
[Quantum dimension of $f \equiv \AAI$] We start from the planar trace of $\id_f$, the identity 2-morphism in $\Hom(\AAI, \AAI)$.  The adjoint of $f$ is $f^* = \IAA$ with $f^* \circ f = \AAI \ot \IAA = \tensor*[_\A]{\A}{} \ot \tensor*{\A}{_\A}$ and $f \circ f^* = \IAA \ot[A] \AAI = \tensor*[_\1]{\A}{_\1}$with quotient map
\begin{align*}
    \cl 0 \cl 0 + \cl 1 \cl 1 &\mapsto \cl 0 \\
    \cl 0 \cl 1 + \cl 1 \cl 0 &\mapsto \cl 1.
\end{align*}
The units and counits are given by
\begin{align*}
&\eta_f:\III \rightarrow \tensor*[_\1]{\A}{_\1} : \cl 0 \mapsto \tau \cl 0\\
&\epsilon_f: \tensor*[_\A]{\A}{} \ot \tensor*{\A}{_\A} \rightarrow \AAA: \cl 0 \cl 0 \mapsto \tau^{-1} \cl 0;\ \cl 1 \cl 1 \mapsto \tau^{-1} \cl 0;\ \cl 0 \cl 1 \mapsto \tau^{-1} \cl 1;\ \cl 1 \cl 0 \mapsto \tau^{-1} \cl 1,
\end{align*}
and
\begin{align*}
    &\eta_{f^\ast}: \AAA \rightarrow \tensor*[_\A]{\A}{} \ot \tensor*{\A}{_\A}: \cl 0 \mapsto \gamma (\cl 0 \cl 0+ \cl 1 \cl 1);\  \cl 1 \mapsto \gamma (\cl 0 \cl 1+ \cl 1 \cl 0),\\
    &\epsilon_{f^\ast}: \tensor*[_\1]{\A}{_\1} \rightarrow \III: \cl 0 \mapsto \gamma^{-1} \cl 0,
\end{align*}
where $\gamma$ and $\tau$ are non-zero complex numbers.  Then the planar traces of $\id_f$ reads
\begin{align*}
&\mathrm{Tr_L}(\id_f):  \III \rightarrow \III: \cl 0 \mapsto \tau \gamma^{-1} \cl 0,\\
&\mathrm{Tr_R}(\id_f): \AAA \rightarrow \AAA: \cl 0 \mapsto 2 \gamma \tau^{-1} \cl 0; \ \cl 1 \mapsto 2 \gamma \tau^{-1} \cl 1.
\end{align*}
Thus the planar trace is in general dependent on the values of $\gamma$ and $\tau$, hence on the choices of the units and counits.  We will show below that the spherical structure imposes extra constraints, which largely reduces the freedom on the choices of units/counits and leads to a more deterministic planar trace.

For $\Sigma \mathrm{sVec}$, both of the objects $\1$ and $\A$ are self-dual with folds $e_\1=i_\1 = \III$ and $e_\A=\VAA$, $i_A=\AAV$ respectively.  According to eqn. \eqref{eq:back-trace}, the back 2-spherical trace of $\id_f$ reads
\begin{align*}
    \mathrm{Tr_B} (\id_f) &= \mathrm{Tr_L} ((\id_f \ot \1) \circ i_{\1})=\mathrm{Tr_L} (\id_f)=\tau \gamma^{-1}, \\
     &=  \mathrm{Tr_R}(e_\A \circ (\id_f \ot \A))= 2\gamma \tau^{-1}.
\end{align*}
Therefore, $2\gamma \tau^{-1}=\tau \gamma^{-1}$, which leads to $\tau \gamma^{-1}=\pm \sqrt{2}$.  In the following, we will choose the units and counits such that the quantum dimensions are positive numbers, hence $\dim(\AAI) = \sqrt{2}$ consistent with the quantum dimension of a Majorana zero mode.
\end{example}

With the same approach, we can compute the quantum dimensions of all the representative 1-morphisms (and their duals), which are all $1$ except that $\dim(\AAI)=\dim(\AAAAA)=\sqrt{2}$ (same for their duals).  We can also calculate the quantum dimension of the objects $\1$ and $\A$, which are given by $\dim(\1) := \dim(\id_\1) = \dim(\III) = 1$ and $\dim(\A) := \dim(\id_\A) = \dim(\AAA) = 1$ respectively.

\section{One example of 10j-symbol in $\Sigma \mathrm{sVec}$}\label{sVec10j}

\begin{figure}
\adjustbox{scale=0.7,center}{
\begin{tikzcd}
\itk{
    \coordinate (a)  at (-3,4);
    \coordinate (b)  at (-1,4);
    \coordinate (c)  at (1,4);
    \coordinate (d)  at (3,4);
    \coordinate (p) at (-2,3);
    \coordinate (x) at (-1,2);
    \coordinate (z)  at (0,1);
    \coordinate (y)  at (1,2);
    \coordinate (q)  at (2,3);
    \coordinate (w)  at (0,3);
    \coordinate (k)  at (0,-1);
    \node[above] at (a) {$ $};
    \node[above] at (b) {$ $};
    \node[above] at (c) {$ $};
    \node[above] at (d) {$ $};
    \node[above] at (p) {$P_3$};
    \node[above] at (x) {$P_2$};
    \node[above] at (z) {$P_1$};
    \node[below] at (k) {$ $};
    \draw [dashed] (a) -- (p);
    \draw [very thick] (p) -- (x);
    \draw [very thick] (x)-- (z);
    \draw [dashed] (z) -- (k);
    \draw [dashed] (b) -- (p);
    \draw [very thick] (c) -- (x);
    \draw [dashed] (d) -- (z);
}
\arrow[rightarrow,d,"","\mid \widetilde{\zeta^1} \rangle"']
\arrow[rightarrow,ddd, bend right=40,"", "\mid {Z} \rangle"'{name=U}]
&&
\itk{
    \coordinate (a)  at (-3,4);
    \coordinate (b)  at (-1,4);
    \coordinate (c)  at (1,4);
    \coordinate (d)  at (3,4);
    \coordinate (p) at (-2,3);
    \coordinate (x) at (-1,2);
    \coordinate (z)  at (0,1);
    \coordinate (y)  at (1,2);
    \coordinate (q)  at (2,3);
    \coordinate (w)  at (0,3);
    \coordinate (k)  at (0,-1);
    \node[above] at (a) {$ $};
    \node[above] at (b) {$ $};
    \node[above] at (c) {$ $};
    \node[above] at (d) {$ $};
    \node[above] at (p) {$P_3$};
    \node[above] at (x) {$P_2$};
    \node[above] at (z) {$P_1$};
    \node[below] at (k) {$ $};
    \draw [dashed] (a) -- (p);
    \draw [very thick] (p) -- (x);
    \draw [very thick] (x)-- (z);
    \draw [dashed] (z) -- (k);
    \draw [dashed] (b) -- (p);
    \draw [very thick] (c) -- (x);
    \draw [dashed] (d) -- (z);
}
\arrow[d,rightarrow,"\mid \widetilde{\zeta^3} \rangle"]
\arrow[rightarrow,ddd, bend left=40,"\mid YWXJ \rangle"{name=N}]
\arrow[rightarrow,ll,"\beta"]
\\
\itk{
    \coordinate (a)  at (-3,4);
    \coordinate (b)  at (-1,4);
    \coordinate (c)  at (1,4);
    \coordinate (d)  at (3,4);
    \coordinate (p) at (-2,3);
    \coordinate (x) at (-1,2);
    \coordinate (z)  at (0,1);
    \coordinate (y)  at (1,2);
    \coordinate (q)  at (2,3);
    \coordinate (w)  at (0,3);
    \coordinate (k)  at (0,-1);
    \node[above] at (a) {$ $};
    \node[above] at (b) {$ $};
    \node[above] at (c) {$ $};
    \node[above] at (d) {$ $};
    \node[above] at (p) {$P_3$};
    \node[above left] at (y) {$Q_3$};
    \node[red, left] at (z) {$Z$};
    \node[below] at (k) {$ $};
    \draw [dashed] (a) -- (p);
    \draw [very thick] (p) -- (z);
    \draw [dashed] (z) -- (k);
    \draw [dashed] (b) -- (p);
    \draw [very thick] (c) -- (y);
    \draw [dashed] (d) -- (y);
    \draw [dashed] (y) -- (z);
} \arrow[rightarrow,d,"","\text{Ic}"']
&&
\itk{
    \coordinate (a)  at (-3,4);
    \coordinate (b)  at (-1,4);
    \coordinate (c)  at (1,4);
    \coordinate (d)  at (3,4);
    \coordinate (p) at (-2,3);
    \coordinate (x) at (-1,2);
    \coordinate (z)  at (0,1);
    \coordinate (y)  at (1,2);
    \coordinate (q)  at (2,3);
    \coordinate (w)  at (0,3);
    \coordinate (k)  at (0,-1);
    \node[above] at (a) {$ $};
    \node[above] at (b) {$ $};
    \node[above] at (c) {$ $};
    \node[above] at (d) {$ $};
    \node[red,above] at (w) {$W$};
    \node[red,above] at (x) {$X$};
    \node[above] at (z) {$P_1$};
    \node[below] at (k) {$ $};
    \draw[dashed] (a) -- (x);
    \draw[very thick] (x)  -- (z);
    \draw[dashed] (b) -- (w);
    \draw[very thick] (c) -- (w);
    \draw[dashed] (z) -- (k);
    \draw[red, very thick] (w) --node[red,right]{$J$} (x);
    \draw[dashed] (d) -- (z);
}\arrow[d,rightarrow,"\mid \widetilde{\zeta^4} \rangle"]
\\
\itk{
    \coordinate (a)  at (-3,4);
    \coordinate (b)  at (-1,4);
    \coordinate (c)  at (1,4);
    \coordinate (d)  at (3,4);
    \coordinate (p) at (-2,3);
    \coordinate (x) at (-1,2);
    \coordinate (z)  at (0,1);
    \coordinate (y)  at (1,2);
    \coordinate (q)  at (2,3);
    \coordinate (w)  at (0,3);
    \coordinate (k)  at (0,-1);
    \node[above] at (a) {$ $};
    \node[above] at (b) {$ $};
    \node[above] at (c) {$ $};
    \node[above] at (d) {$ $};
    \node[above right] at (x) {$P_3$};
    \node[above] at (q) {$Q_3$};
    \node[red,left] at (z) {$Z$};
    \node[below] at (k) {$ $};
    \draw [dashed](a) -- (x);
    \draw [very thick](x) -- (z);
    \draw [dashed](z) -- (k);
    \draw [dashed](b) -- (x);
    \draw [very thick](c) -- (q);
    \draw [dashed](d) -- (q) -- (z);
} \arrow[rightarrow,d,"","\mid \widetilde{\zeta^2}\rangle"']
&&
\itk{
    \coordinate (a)  at (-3,4);
    \coordinate (b)  at (-1,4);
    \coordinate (c)  at (1,4);
    \coordinate (d)  at (3,4);
    \coordinate (p) at (-2,3);
    \coordinate (x) at (-1,2);
    \coordinate (z)  at (0,1);
    \coordinate (y)  at (1,2);
    \coordinate (q)  at (2,3);
    \coordinate (w)  at (0,3);
    \coordinate (k)  at (0,-1);
    \node[above] at (a) {$ $};
    \node[above] at (b) {$ $};
    \node[above] at (c) {$ $};
    \node[above] at (d) {$ $};
    \node[red,above] at (w) {$W$};
    \node[red,above] at (y) {$Y$};
    \node[above] at (z) {$Q_1$};
    \node[below] at (k) {$ $};
    \draw [dashed] (z) -- (k);
    \draw [dashed](a) -- (z);
    \draw [dashed](b) -- (w);
    \draw [very thick](c) -- (w);
    \draw [red, very thick](w) --node[red,left]{$J$} (y);
    \draw [dashed](d) -- (y);
    \draw [very thick](y) -- (z);
}
\arrow[d,rightarrow,"\mid \widetilde{\zeta^5}\rangle"]
\\
\itk{
    \coordinate (a)  at (-3,4);
    \coordinate (b)  at (-1,4);
    \coordinate (c)  at (1,4);
    \coordinate (d)  at (3,4);
    \coordinate (p) at (-2,3);
    \coordinate (x) at (-1,2);
    \coordinate (z)  at (0,1);
    \coordinate (y)  at (1,2);
    \coordinate (q)  at (2,3);
    \coordinate (w)  at (0,3);
    \coordinate (k)  at (0,-1);
    \node[above] at (a) {$ $};
    \node[above] at (b) {$ $};
    \node[above] at (c) {$ $};
    \node[above] at (d) {$ $};
    \node[above] at (y) {$Q_2$};
    \node[above] at (q) {$Q_3$};
    \node[above] at (z) {$Q_1$};
    \node[below] at (k) {$ $};
    \draw [dashed] (z) -- (k);
    \draw [dashed](a) -- (z);
    \draw [dashed](b) -- (y);
    \draw [very thick](c) -- (q);
    \draw [very thick](y) -- (z);
    \draw [dashed](d) -- (q) -- (y);
}
\arrow[rightarrow,rr,"g"]
&&
\itk{
    \coordinate (a)  at (-3,4);
    \coordinate (b)  at (-1,4);
    \coordinate (c)  at (1,4);
    \coordinate (d)  at (3,4);
    \coordinate (p) at (-2,3);
    \coordinate (x) at (-1,2);
    \coordinate (z)  at (0,1);
    \coordinate (y)  at (1,2);
    \coordinate (q)  at (2,3);
    \coordinate (w)  at (0,3);
    \coordinate (k)  at (0,-1);
    \node[above] at (a) {$ $};
    \node[above] at (b) {$ $};
    \node[above] at (c) {$ $};
    \node[above] at (d) {$ $};
    \node[above] at (y) {$Q_2$};
    \node[above] at (q) {$Q_3$};
    \node[above] at (z) {$Q_1$};
    \node[below] at (k) {$ $};
    \draw [dashed] (z) -- (k);
    \draw [dashed](a) -- (z);
    \draw [dashed](b) -- (y);
    \draw [very thick](c) -- (q);
    \draw [very thick](y) -- (z);
    \draw [dashed](d) -- (q) -- (y);
}
\end{tikzcd}}
\caption{An example of 10j-symbol. Dashed lines represent object $\one$ and solid lines represent object $\A$. Red line and points represent object and 1-morphisms which are variables.}
\label{10jexample}
\end{figure}
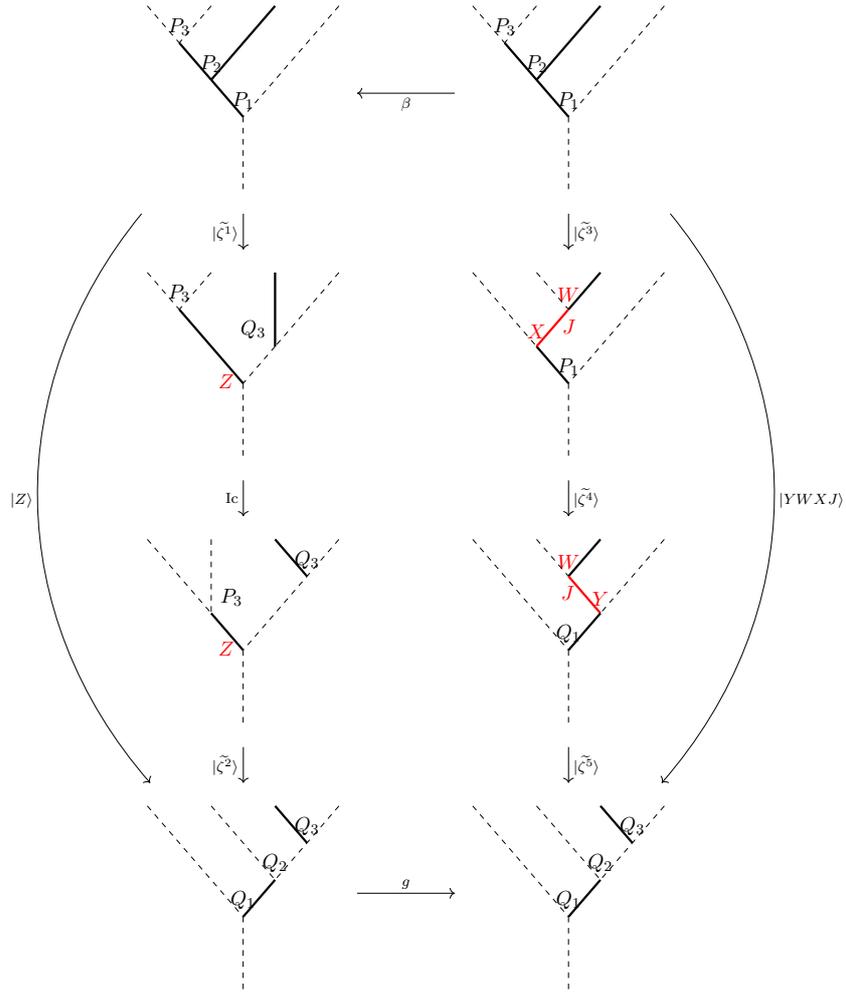

In this section, we will show how to calculate $G$ and $G^{-1}$ for $P_1 = Q_1 = Q_3 = \AAI$,  $P_2 = \AAAAA$, and $P_3 = Q_2 = \IAA$, which has been depicted as Fig.\ref{10jexample}.  In the figure, the dashed and solid lines correspond to the object $\1$ and $\A$ respectively.  For readers who want to skip the technical details, the results of this example can be found in eqn. \eqref{explicit10j}.

The 10j-symbol is just the transformation between the two bases $\cl{Z}$ and $\cl{YWXJ}$.  We start from $\cl{Z}$.  Recall that $YWXZ$ can only be chosen from the the representative 1-morphisms, which are $\III$, $\IfI$, $\IAA$, $\AAI$, $\AAA$, $\AfAA$, $\AAV$, $\AAVf$, and $\AAAAA$, hence we have $Z = \AAI$.  The retraction ${\zeta^1}$ is given by the direct sum decomposition $(P_2 \ot \1) \circ P_1 = \bigoplus_Z (A \ot Q_3) \circ Z$.  And we have $(P_2 \ot \1) \circ P_1 = \AAAAA \circ \AAI = \AAAAI \xrightarrow[\sim]{p_1} \AAV \oplus \AAVf$ and $(\A \ot Q_3) \circ Z = (\A \ot \AAI) \circ \AAI = \tensor*[_{\A \ot \A}]{}{} \widetilde{\A \ot \A} \tensor*{}{_\1} \xrightarrow[\sim]{p_2} \AAV \oplus \AAVf$, where $\widetilde{\A \ot \A}$ is same vector space as $\A \ot \A$, but with different actions. $p_1$ and $p_2$ is given by
\begin{align*}
p_1&=\frac{1}{\sqrt{2}}\bordermatrix{~ & \cl {00} & \cl {01} & \cl {10} & \cl {11} \cr
              \cl {00}_V & 1 & 0 & 0 & i \cr
              \cl {01}_{Vf} & 0 & 1 & i & 0 \cr
              \cl {10}_V & 0 & 1 & -i & 0 \cr
              \cl {11}_{Vf} & 1 & 0 & 0 & -i \cr} &
    p_2&=\frac{1}{\sqrt{2}}\bordermatrix{~ &  \cl {00} & \cl {01} & \cl {10} & \cl {11} \cr
              \cl {00}_V & 1 & 0 & 0 & -i \cr
              \cl {01}_{Vf} & 0 & i & 1 & 0 \cr
              \cl {10}_V & 0 & -i & 1 & 0 \cr
              \cl {11}_{Vf} & 1 & 0 & 0 & i \cr}.
\end{align*}

It is obvious that the retraction map in the direct sum composition is given by ${\zeta^1} = p_2^{-1} \cdot \id_{V \oplus V_f} \cdot p_1 = p_2^{-1} \cdot \id_{V} \cdot p_1 + p_2^{-1} \cdot \id_{V_f} \cdot p_1 \equiv {\zeta^1}_0 + {\zeta^1}_1$ with
\begin{align*}
{\zeta^1}_0&=\frac{1}{2}\bordermatrix{~ & \cl{00} & \cl{01} & \cl{10} & \cl{11} \cr
              \cl{00} & 1 & 0 & 0 & i \cr
              \cl{01} & 0 & i & 1 & 0 \cr
              \cl{10} & 0 & 1 & -i & 0 \cr
              \cl{11} & i & 0 & 0 & -1 \cr}&
{\zeta^1}_1&=\frac{1}{2}\bordermatrix{~ & \cl{00} & \cl{01} & \cl{10} & \cl{11} \cr
              \cl{00} & 1 & 0 & 0 & -i \cr
              \cl{01} & 0 & -i & 1 & 0 \cr
              \cl{10} & 0 & 1 & i & 0 \cr
              \cl{11} & -i & 0 & 0 & -1 \cr},
\end{align*}
where the subscript $0$ suggests the bimodule map is between representative simple 1-morphisms with no decorated fermion, for example, $\III$, $\AAA$ and $\AAV$, while the subscript $1$ suggests that the bimodule map is between representative simple 1-morphisms with a (decorated) fermion, for example, $\IfI$, $\AfAA$ and $\AAVf$.
Since the normalization of ${\zeta^1}$ is trivial, we have
\begin{align*}
    \cl {\zeta^1} = {\zeta^1}=\bordermatrix{~ & \cl{00} & \cl{01} & \cl{10} & \cl{11} \cr
              \cl{00} & 1 & 0 & 0 & 0 \cr
              \cl{01} & 0 & 0 & 1 & 0 \cr
              \cl{10} & 0 & 1 & 0 & 0 \cr
              \cl{11} & 0 & 0 & 0 & -1 \cr}.
\end{align*}

In $\Sigma \mathrm{sVec}$, the associator of tensor product $\alpha$ and the associator bimodules $\Lambda$ are all trivial. For $\cl {\tilde{\zeta}^1}$, the associator of bimodule compositions $\lambda$ is also trivial. The calculation of $\mid \tilde{\zeta}^1 \rangle$ is vastly simplified and given by the following diagram
    \begin{equation}
        \begin{tikzcd}
        	{(\A\ot \A)\ot (\A\ot \A)} && {(\A\ot \A)\ot \widetilde{(\A\ot \A)}} \\
        	\\
        	{(\A\ot \A) \circ (\A\ot \A)} = (\A\ot \A) && {(\A\ot \A) \circ \widetilde{(\A\ot \A)}} = \widetilde{(\A\ot \A)}
        	\arrow[rightarrow,"\id_{(\A\ot \A)} \ot \mid {\zeta^1} \rangle ", from=1-1, to=1-3]
        	\arrow["\pi_1"', from=1-1, to=3-1]
        	\arrow[rightarrow,"\mid \tilde{\zeta}^1 \rangle"', from=3-1, to=3-3]
        	\arrow["\pi_2", from=1-3, to=3-3]
        \end{tikzcd},
        \label{aanda'}
    \end{equation}
where $\pi_1$ and $\pi_2$ are the quotient map in the relative tensor product.  With the standard protocol for selecting basis, the matrix of $\pi_1$ and $\pi_2$ are same, hence the matrix of $\cl {\tilde{\zeta}^1_0}$ ($\cl {\tilde{\zeta}^1_1}$) is same as the matrix of ${\zeta^1}_0$ (${\zeta^1}_1$).

Similarly, we can get the interchange bimodule map
\begin{align*}
\mathrm{Ic}=\bordermatrix{~ & \cl{00} & \cl{01} & \cl{10} & \cl{11} \cr
              \cl{00} & 1 & 0 & 0 & 0 \cr
              \cl{01} & 0 & 1 & 0 & 0 \cr
              \cl{10} & 0 & 0 & 1 & 0 \cr
              \cl{11} & 0 & 0 & 0 & 1 \cr},
\end{align*}
and the normalized retraction bimodule maps
\begin{align*}
\cl {\tilde{\zeta}^2_0}&=\bordermatrix{~ & \cl{00} & \cl{01} & \cl{10} & \cl{11} \cr
              \cl{00} & 1 & 0 & 0 & 0 \cr
              \cl{01} & 0 & 0 & 0 & 0 \cr
              \cl{10} & 0 & 0 & 1 & 0 \cr
              \cl{11} & 0 & 0 & 0 & 0 \cr} &
\cl {\tilde{\zeta}^2_1}&=\bordermatrix{~ & \cl{00} & \cl{01} & \cl{10} & \cl{11} \cr
              \cl{00} & 0 & 0 & 0 & 0 \cr
              \cl{01} & 0 & 1 & 0 & 0 \cr
              \cl{10} & 0 & 0 & 0 & 0 \cr
              \cl{11} & 0 & 0 & 0 & 1 \cr}.
\end{align*}

By composing $\cl {\tilde{\zeta}^1}$, $\mathrm{Ic}$ and $\cl{\tilde{\zeta}^2}$, we have$\cl Z = \oplus_{a,b} \mid Z_{ab} \rangle =\oplus_{a,b} \cl{\tilde{\zeta}_b^2} \cdot \mathrm{Ic} \cdot \cl {\tilde{\zeta}_a^1}$ with
\begin{align*}
\mid Z_{00} \rangle&=\frac{1}{2}\bordermatrix{~ & \cl{00} & \cl{01} & \cl{10} & \cl{11} \cr
              \cl{00} & 1 & 0 & 0 & i \cr
              \cl{01} & 0 & 0 & 0 & 0 \cr
              \cl{10} & 0 & 1 & -i & 0 \cr
              \cl{11} & 0 & 0 & 0 & 0 \cr}, &
\mid Z_{01} \rangle&=\frac{1}{2}\bordermatrix{~ & \cl{00} & \cl{01} & \cl{10} & \cl{11} \cr
              \cl{00} & 0 & 0 & 0 & 0 \cr
              \cl{01} & 0 & i & 1 & 0 \cr
              \cl{10} & 0 & 0 & 0 & 0 \cr
              \cl{11} & i & 0 & 0 & -1 \cr},\\
\mid Z_{10} \rangle&=\frac{1}{2}\bordermatrix{~ & \cl{00} & \cl{01} & \cl{10} & \cl{11} \cr
              \cl{00} & 1 & 0 & 0 & -i \cr
              \cl{01} & 0 & 0 & 0 & 0 \cr
              \cl{10} & 0 & 1 & i & 0 \cr
              \cl{11} & 0 & 0 & 0 & 0 \cr},&
\mid Z_{11} \rangle&=\frac{1}{2}\bordermatrix{~ & \cl{00} & \cl{01} & \cl{10} & \cl{11} \cr
              \cl{00} & 0 & 0 & 0 & 0 \cr
              \cl{01} & 0 & -i & 1 & 0 \cr
              \cl{10} & 0 & 0 & 0 & 0 \cr
              \cl{11} & -i & 0 & 0 & -1 \cr}.
\end{align*}

For $\cl{YWXJ}$, in the case of $\Sigma \mathrm{sVec}$, the object $J$ is uniquely determined by $Y$, $W$, and $X$, hence it reduce to $\cl {YWX}$, which has in total nine different choices.  Here we show the result with $Y=W=\AfAA$ and $X=\AAA$ as an example, where $\mid YWX \rangle$ reads
\begin{align*}
 \mid YWX \rangle=\frac{1}{2}\bordermatrix{~ & \cl{00} & \cl{01} & \cl{10} & \cl{11} \cr
               \cl{00} & 0 & 0 & 0 & -1 \cr
               \cl{01} & 0 & 0 & 1 & 0 \cr
               \cl{10} & 0 & 0 & 1 & 0 \cr
               \cl{11} & 0 & 0 & 0 & -1 \cr},
\end{align*}
Therefore, we have
\begin{align}
\label{eq:YWXZ}
    \mid YWX \rangle=\frac{i}{2}(\mid Z_{00}\rangle-\mid Z_{10}\rangle)+\frac{1}{2}(\mid Z_{01}\rangle+\mid Z_{11}\rangle)
\end{align}


For better presentation of the 10j-symbols, we divide the representative 1-morphisms into three groups
\begin{itemize}
    \item the bimodules between Morita non-equivalent objects such as $\IAA, \tensor*[_\A]{f\A}{_\1}, \AAAAA$ etc, which are denoted as $\mu$
    \item the bimodules between Morita equivalent objects and decorated by one fermion, for example, $\IfI$, $\AfAA$, $\AAVf$ etc, which are denoted as $f$.
    \item the bimodules between Morita equivalent objects with no fermion decoration, for example, $\III$, $\AAA$, $\AAV$ etc, which are denoted as $1$.
\end{itemize}
Then eqn. \eqref{eq:YWXZ} becomes $\mid ff1 \rangle=\frac{i}{2}(\mid {\mu}_{00}\rangle-\mid {\mu}_{10}\rangle)+\frac{1}{2}(\mid {\mu}_{01}\rangle+\mid {\mu}_{11}\rangle)$ or matrix elements
\begin{align*}
    G^{ff1}_{\mu_{00}} &= i/2 & G^{ff1}_{\mu_{01}} &= 1/2 & G^{ff1}_{\mu_{10}} &= -i/2 & G^{ff1}_{\mu_{11}} &= 1/2
\end{align*}

The final results for the Fig.\ref{10jexample} are
\begin{align}
G^{T}=\frac{1}{\sqrt{2}}
\bordermatrix{~ & {\mu}_{00} & {\mu}_{01} & {\mu}_{10} & {\mu}_{11} \cr
              {\mu\mu\mu}_{000} & \frac{1}{\sqrt{2}} & -\frac{i}{\sqrt{2}} & \frac{1}{\sqrt{2}} & \frac{i}{\sqrt{2}}\cr
              {\mu\mu\mu}_{001} & \frac{1}{\sqrt{2}} & \frac{i}{\sqrt{2}} & \frac{1}{\sqrt{2}} & -\frac{i}{\sqrt{2}}\cr
              {\mu\mu\mu}_{010} & \frac{1}{\sqrt{2}} & -\frac{i}{\sqrt{2}} & \frac{1}{\sqrt{2}} & \frac{i}{\sqrt{2}}\cr
              {\mu\mu\mu}_{011} & -\frac{1}{\sqrt{2}} & -\frac{i}{\sqrt{2}} & -\frac{1}{\sqrt{2}} & \frac{i}{\sqrt{2}}\cr
              {\mu\mu\mu}_{100} & -\frac{i}{\sqrt{2}} & -\frac{1}{\sqrt{2}} & \frac{i}{\sqrt{2}} & -\frac{1}{\sqrt{2}}\cr
              {\mu\mu\mu}_{101} & -\frac{i}{\sqrt{2}} & \frac{1}{\sqrt{2}} & \frac{i}{\sqrt{2}} & \frac{1}{\sqrt{2}}\cr
              {\mu\mu\mu}_{110} & \frac{i}{\sqrt{2}} & \frac{1}{\sqrt{2}} & -\frac{i}{\sqrt{2}} & \frac{1}{\sqrt{2}}\cr
              {\mu\mu\mu}_{111} & -\frac{i}{\sqrt{2}} & \frac{1}{\sqrt{2}} & \frac{i}{\sqrt{2}} & \frac{1}{\sqrt{2}}\cr
              ff1 & \frac{i}{\sqrt{2}} & \frac{1}{\sqrt{2}} & -\frac{i}{\sqrt{2}} & \frac{1}{\sqrt{2}}\cr
              fff & \frac{1}{\sqrt{2}} & -\frac{i}{\sqrt{2}} & \frac{1}{\sqrt{2}} & \frac{i}{\sqrt{2}}\cr
              f11 & \frac{1}{\sqrt{2}} & \frac{i}{\sqrt{2}} & \frac{1}{\sqrt{2}} & -\frac{i}{\sqrt{2}}\cr
              f1f & -\frac{i}{\sqrt{2}} & \frac{1}{\sqrt{2}} & \frac{i}{\sqrt{2}} & \frac{1}{\sqrt{2}}\cr
              1f1 & -\frac{i}{\sqrt{2}} & \frac{1}{\sqrt{2}} & \frac{i}{\sqrt{2}} & \frac{1}{\sqrt{2}}\cr
              1ff & -\frac{1}{\sqrt{2}} & -\frac{i}{\sqrt{2}} & -\frac{1}{\sqrt{2}} & \frac{i}{\sqrt{2}}\cr
              111 & \frac{1}{\sqrt{2}} & -\frac{i}{\sqrt{2}} & \frac{1}{\sqrt{2}} & \frac{i}{\sqrt{2}}\cr
              11f & -\frac{i}{\sqrt{2}} & -\frac{1}{\sqrt{2}} & \frac{i}{\sqrt{2}} & -\frac{1}{\sqrt{2}}\cr},
G^{-1}=\frac{1}{4\sqrt{2}}
\bordermatrix{~ & {\mu}_{00} & {\mu}_{01} & {\mu}_{10} & {\mu}_{11} \cr
              {\mu\mu\mu}_{000} & \frac{1}{\sqrt{2}} & \frac{i}{\sqrt{2}} & \frac{1}{\sqrt{2}} & -\frac{i}{\sqrt{2}}\cr
              {\mu\mu\mu}_{001} & \frac{1}{\sqrt{2}} & -\frac{i}{\sqrt{2}} & \frac{1}{\sqrt{2}} & \frac{i}{\sqrt{2}}\cr
              {\mu\mu\mu}_{010} & \frac{1}{\sqrt{2}} & \frac{i}{\sqrt{2}} & \frac{1}{\sqrt{2}} & -\frac{i}{\sqrt{2}}\cr
              {\mu\mu\mu}_{011} & -\frac{1}{\sqrt{2}} & \frac{i}{\sqrt{2}} & -\frac{1}{\sqrt{2}} & -\frac{i}{\sqrt{2}}\cr
              {\mu\mu\mu}_{100} & \frac{i}{\sqrt{2}} & -\frac{1}{\sqrt{2}} & -\frac{i}{\sqrt{2}} & -\frac{1}{\sqrt{2}}\cr
              {\mu\mu\mu}_{101} & \frac{i}{\sqrt{2}} & \frac{1}{\sqrt{2}} & -\frac{i}{\sqrt{2}} & \frac{1}{\sqrt{2}}\cr
              {\mu\mu\mu}_{110} & -\frac{i}{\sqrt{2}} & \frac{1}{\sqrt{2}} & \frac{i}{\sqrt{2}} & \frac{1}{\sqrt{2}}\cr
              {\mu\mu\mu}_{111} & \frac{i}{\sqrt{2}} & \frac{1}{\sqrt{2}} & -\frac{i}{\sqrt{2}} & \frac{1}{\sqrt{2}}\cr
              ff1 & -\frac{i}{\sqrt{2}} & \frac{1}{\sqrt{2}} & \frac{i}{\sqrt{2}} & \frac{1}{\sqrt{2}}\cr
              fff & \frac{1}{\sqrt{2}} & \frac{i}{\sqrt{2}} & \frac{1}{\sqrt{2}} & -\frac{i}{\sqrt{2}}\cr
              f11 & \frac{1}{\sqrt{2}} & -\frac{i}{\sqrt{2}} & \frac{1}{\sqrt{2}} & \frac{i}{\sqrt{2}}\cr
              f1f & \frac{i}{\sqrt{2}} & \frac{1}{\sqrt{2}} & -\frac{i}{\sqrt{2}} & \frac{1}{\sqrt{2}}\cr
              1f1 & \frac{i}{\sqrt{2}} & \frac{1}{\sqrt{2}} & -\frac{i}{\sqrt{2}} & \frac{1}{\sqrt{2}}\cr
              1ff & -\frac{1}{\sqrt{2}} & \frac{i}{\sqrt{2}} & -\frac{1}{\sqrt{2}} & -\frac{i}{\sqrt{2}}\cr
              111 & \frac{1}{\sqrt{2}} & \frac{i}{\sqrt{2}} & \frac{1}{\sqrt{2}} & -\frac{i}{\sqrt{2}}\cr
              11f &\frac{i}{\sqrt{2}} & -\frac{1}{\sqrt{2}} & -\frac{i}{\sqrt{2}} & -\frac{1}{\sqrt{2}}\cr}.
              \label{explicit10j}
\end{align}
One can easily check that $GG^{-1}=1$.

\section{Conclusion}
In conclusion, we propose a method to construct a class of fusion 2-category $\Sigma \cB$ and obtain all its categorical data. We apply this method to $\Sigma \mathrm{sVec}$ to work out all its categorical data explicitly. All the 10j-symbols of $\Sigma \mathrm{sVec}$ and the complete 
computer program has been uploaded to github. With the example, we demonstrate that our method can be efficiently encoded to calculate all wanted categorical data in computer program.

\acknowledgments
We are grateful to the helpful discussions with Thibault Décoppet. This work was supported by the National Key R$\&$D Program of China (Grants No. 2022YFA1403700), NSFC (Grants No. 12141402), the Science, Technology and Innovation Commission of Shenzhen Municipality (No. ZDSYS20190902092905285), and Center for Computational Science and Engineering at Southern University of Science and Technology. TL is supported by start-up funding from The Chinese University of Hong Kong. LW and TL are also supported by funding from Hong Kong Research Grants Council
(ECS No.~24304722). WX and CW are supported by Research Grants Council of Hong Kong (GRF 17311322) and National Natural Science Foundation of China (Grant No. 12222416).

\appendix
\section{Direct sum decomposition of $\A \ot \A$ as a left ${\A\otimes \A}$-module}
\label{sec:decom_AA}
$\A\otimes \A$ can be regarded as a left ${\A\otimes \A}$-module with the left action
\begin{align}
\cl a\cl b \act \cl c\cl d:=\cl a\cl b \bullet \cl c\cl d:= (-)^{bc} \cl {a+c} \cl {b+d}.
\end{align}
We start from a vector
\[ \cl 0\cl0+\alpha \cl 1\cl1,\]
under the $\A \ot \A$-action, we have
\begin{align*}
\cl0\cl1\act (\cl 0\cl0+\alpha \cl 1\cl1)&=\cl0\cl1-\alpha\cl1\cl0,\\
\cl1\cl0\act (\cl 0\cl0+\alpha \cl 1\cl1)&=\cl1\cl0+\alpha\cl0\cl1=\alpha(\cl0\cl1+\alpha^{-1} \cl1\cl0),\\
\cl1\cl1\act (\cl 0\cl0+\alpha \cl 1\cl1)&=\cl1\cl1-\alpha\cl0\cl0=-\alpha(\cl 0\cl0-\alpha^{-1} \cl 1\cl1).
\end{align*}
We found that it is closed if we choose $\alpha=-\alpha^{-1}$, namely $\alpha=\pm \ii$, which gives the direct sum decomposition
\begin{align*}
    \A \ot \A = \tilde{V} \oplus \tilde{V}',
\end{align*}
with $\tilde{V} = \mathrm{Span}\{\cl0 \cl0+\ii\cl1\cl1, \cl0\cl1-\ii\cl1\cl0\}$ and $\tilde{V}'= \mathrm{Span}\{\cl0 \cl0-\ii\cl1\cl1, \cl0\cl1+\ii\cl1\cl0\}$.  It can be easily show that $\tilde{V} \cong V$ and $\tilde{V'} \cong V'$ in example \ref{ex:left_mod_AA}.

\section{Relative tensor product of $V^{rev} \ot[\A \ot \A] V$}
\label{sec:VVTI}

The bases of $V^{rev}$, $\A\ot \A$, and $V$ are denoted as $\cl m$, $\cl {ab}$, and $\cl n$, with $a, b, m, n \in \{0, 1\}$, respectively. The right $\A\ot \A$-action on $V^{rev}$ is given by
\begin{align}
\cl m\ract\cl a\cl b=(-)^{(a+m)b}(\ii)^b\cl{a+b+m},
\end{align}
while the left $\A\ot \A$-action on $V$ reads
\begin{align}
\cl a\cl b \act \cl n = (-)^{(b+n)a} (\ii)^a\cl{a+b+n}.
\end{align}
Then, $\cl m \cl n$ form a basis of $V^{rev} \ot V$ where $\cl m$ and $\cl n$ are bases of $V^{rev}$ and $V$ respectively.  The relative tensor product $V^{rev} \ot[\A \ot \A] V$
can be regarded as a subspace of $V^{rev} \ot V$, with a quotient map $V^{rev} \ot V \rightarrow V^{rev} \ot[\A \ot \A] V$ satisfies
\begin{align}
\label{eq:mabn}
    \cl m (\cl {ab}  \act \cl n) - (\cl m\ract \cl {ab} ) \cl n \mapsto 0,
\end{align}
$\forall a, b, m, n \in \{0, 1\}$. 

Some nontrivial ones from eqn. \eqref{eq:mabn} are given below
\begin{align*}
\cl 0 (\cl 0 \cl 1 \act \cl 0) - (\cl 0 \ract \cl 0 \cl 1 ) \cl 0=\cl 0\cl 1  - i \cl 1 \cl 0 \mapsto 0,\\
\cl 0(\cl 1 \cl 0 \act \cl 0) - (\cl 0 \ract \cl 1 \cl 0) \cl 0= i \cl 0 \cl 1 - \cl 1 \cl 0 \mapsto 0, \\
\cl 0 (\cl 0 \cl 1 \act \cl 1) - (\cl 0 \ract \cl 0 \cl 1 ) \cl 1=\cl 0 \cl 0 - i \cl 1 \cl 1 \mapsto 0,
\end{align*}
where the first two leads to $\cl 0\cl 1 \mapsto 0$, and $\cl 0\cl 1 \mapsto 0$. It is obvious that we can choose $V^{rev} \ot[\A \ot \A] V = \mathrm{span}\{\cl 0\cl 0+i \cl 1\cl 1 \}$.  Since it can be easily shown that $\mathrm{span}\{\cl 0\cl 0+i \cl 1\cl 1 \} \cong \III$, we finally have
\begin{align*}
    V^{rev} \ot[\A \ot \A] V & = \III,
\end{align*}
with quotient map 
\begin{align*}
    V^{rev} \ot V &\rightarrow \III \\
    \cl 0\cl 0+i \cl 1\cl 1 &\mapsto \cl 0 \\
    \cl 0\cl 1 &\mapsto 0 \\
    \cl 0\cl 1 &\mapsto 0 \\
    \cl 0 \cl 0 - i \cl 1 \cl 1 &\mapsto 0
\end{align*}

\section{10j-symbols of $\Sigma \mathrm{sVec}$} 
The complete program which can compute all the 10j-symbols of $\Sigma \mathrm{sVec}$ has been uploaded to github. We have also verified the coherence condition\cite{Gur2006} of all these 10j-symbols. For more details, please see https://github.com/WJXI/2sVec.git.

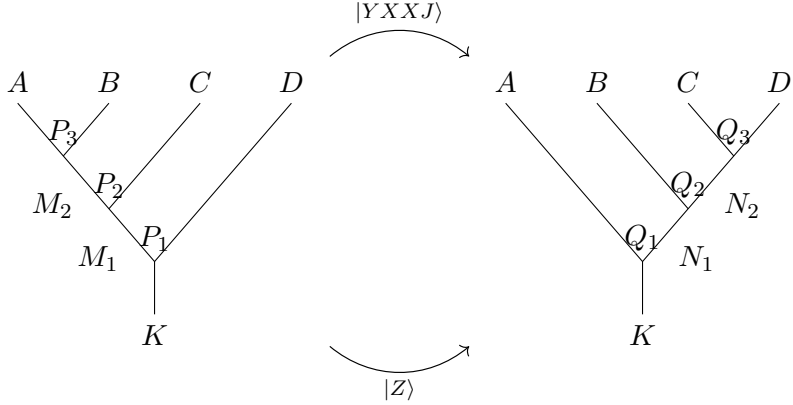
\begin{figure}
\begin{tikzcd}
\itk{
    \coordinate (a)  at (-3,4);
    \coordinate (b)  at (-1,4);
    \coordinate (c)  at (1,4);
    \coordinate (d)  at (3,4);
    \coordinate (p) at (-2,3);
    \coordinate (x) at (-1,2);
    \coordinate (z)  at (0,1);
    \coordinate (y)  at (1,2);
    \coordinate (q)  at (2,3);
    \coordinate (w)  at (0,3);
    \coordinate (k)  at (0,0);
    \node[above] at (a) {$A$};
    \node[above] at (b) {$B$};
    \node[above] at (c) {$C$};
    \node[above] at (d) {$D$};
    \node[above] at (p) {$P_3$};
    \node[above] at (x) {$P_2$};
    \node[above] at (z) {$P_1$};
    \node[below] at (k) {$K$};
    \draw (a) -- (p) --node[below left]{$M_2$} (x) --node[below left]{$M_1$} (z) -- (k);
    \draw (b) -- (p);
    \draw (c) -- (x);
    \draw (d) -- (z);
}
\arrow[rr,bend right=40,"", "\mid Z \rangle"'{name=U}]
\arrow[rr,bend left=40,"\mid YXXJ \rangle", ""'{name=N}]
&&
\itk{
    \coordinate (a)  at (-3,4);
    \coordinate (b)  at (-1,4);
    \coordinate (c)  at (1,4);
    \coordinate (d)  at (3,4);
    \coordinate (p) at (-2,3);
    \coordinate (x) at (-1,2);
    \coordinate (z)  at (0,1);
    \coordinate (y)  at (1,2);
    \coordinate (q)  at (2,3);
    \coordinate (w)  at (0,3);
    \coordinate (k)  at (0,0);
    \node[above] at (a) {$A$};
    \node[above] at (b) {$B$};
    \node[above] at (c) {$C$};
    \node[above] at (d) {$D$};
    \node[above] at (y) {$Q_2$};
    \node[above] at (q) {$Q_3$};
    \node[above] at (z) {$Q_1$};
    \node[below] at (k) {$K$};
    \draw (a) -- (z) -- (k);
    \draw (b) -- (y);
    \draw (c) -- (q);
    \draw (d) -- (q) --node[below right]{$N_2$} (y) --node[below right]{$N_1$} (z);
} 
\end{tikzcd} 
\caption{The initial state(left) and the final state(right) of two maps $\cl Z$ and $\cl {YWXJ}$.}
\label{ifstate}
\end{figure}
The main code is \emph{tjmatrix.m}. Input are the initial state and final state of two maps $\cl Z$ and $\cl {YWXJ}$ and output is a matrix of 10j-symbols which characterizes basis transformation between $\cl Z$ and $\cl {YWXJ}$. As shown in Fig.\ref{ifstate}, the input contains 9 objects and 6 1-morphisms. These 9 objects are $A$, $M_2$, $M_1$, $K$, $B$, $N_1$, $C$, $N_2$ and $D$. (They are placed on 1-simplexes $01$, $02$, $03$, $04$, $12$, $14$, $23$, $24$ and $34$ respectively). Similarly, the 6 1-morphisms are $P_3$, $Q_1$, $P_2$, $P_1$, $Q_2$ and $Q_3$. (They are placed on 2-simplexes $012$, $014$, $023$, $034$, $124$ and $234$ respectively). In the code, object $\1$ is represented by number $0$ and object $\A$ is represented by number $1$. Similarly, 1-morphisms $1$ and $\mu$ are represented by number $0$ and 1-morphism $f$ is represented by number $1$. 

For example, to generate the matrix in the left hand side of equ.~\eqref{explicit10j}, input is an array of objects [0 1 1 0 0 1 1 0 0] and an array of 1-morphisms [0 0 0 0 0 0].

\bibliography{bib}
\bibliographystyle{JHEP}
\end{document}